\documentclass[sn-nature]{sn-jnl}
\usepackage{graphicx}%
\usepackage{multirow}%
\usepackage{amsmath,amssymb,amsfonts}%
\usepackage{amsthm}%
\usepackage{mathrsfs}%
\usepackage[title]{appendix}%
\usepackage{xcolor}%
\usepackage{textcomp}%
\usepackage{manyfoot}%
\usepackage{booktabs}%
\usepackage{algorithm}%
\usepackage{algorithmicx}%
\usepackage{algpseudocode}%
\usepackage{listings}%
\usepackage{subfig}
\usepackage{lineno}
\usepackage{caption}

\usepackage[resetlabels]{multibib}
\newcites{supp}{References}

\usepackage{comment}
\theoremstyle{thmstyleone}%

\theoremstyle{thmstyletwo}%

\theoremstyle{thmstylethree}%

\usepackage{geometry}
\begin{document}

\title [Article Title]{Fully integrated hybrid multimode-multiwavelength photonic processor with picosecond latency}

\author*[1]{\fnm{Ahmed} \sur{Khaled}}\email{20ak41@queensu.ca}

\author[1]{\fnm{A.} \sur{Aadhi}}\email{aadhi.a@queensu.ca}

\author[2]{\fnm{Chaoran} \sur{Huang}}\email{crhuang@ee.cuhk.edu.hk}

\author[3]{\fnm{Alexander N.} \sur{Tait}}\email{atait@ieee.org}

\author*[1]{\fnm{Bhavin J.} \sur{Shastri}}\email{shastri@ieee.org}

\affil[1]{\orgdiv{Centre for Nanophotonics, Department of Physics, Engineering Physics \& Astronomy}, \orgname{Queen's University}, 
\city{Kingston}, \postcode{K7L 3N6}, \state{Ontario}, \country{Canada}}

\affil[2]{\orgdiv{Department of Electronic Engineering}, \orgname{The Chinese University of Hong Kong}, 
\city{Hong Kong SAR}, \country{China}}

\affil[3]{\orgdiv{Smith Engineering, Electrical and Computer Engineering}, \orgname{Queen's University}, 
\city{Kingston}, \postcode{K7L 3N6}, \state{Ontario}, \country{Canada}}


\abstract{High-speed signal processing is essential for maximizing data throughput in emerging communication applications, like multiple-input multiple-output (MIMO) systems and radio-frequency (RF) interference cancellation. However, as these technologies scale, they increase hardware complexity, computing power demands, and create significant digital signal processing (DSP) challenges. While transistor miniaturization has improved digital electronic processors, they still face physical bottlenecks, limiting computational throughput and increasing DSP latency. Photonic processors present a promising alternative, offering large bandwidth, low loss, parallel processing, and low latency. Yet, scalability in photonic processors remains limited by system integration, device size, and on-chip multiplexing challenges. Here, we introduce a scalable on-chip hybrid multiplexed photonic processor, combining mode-division multiplexing (MDM) and wavelength-division multiplexing (WDM). This marks the first implementation of a monolithically integrated MDM-WDM-compatible processor, featuring mode multiplexers, multimode microring resonators, and multimode balanced photodetectors. Furthermore, we demonstrate real-time unscrambling of 5 Gb/s non-return-to-zero optical MIMO signals and RF phase-shift keying signal unjamming. Our system’s 30 ps processing latency makes it ideal for real-time MIMO and RF applications. Our analysis reveals that hybrid MDM-WDM multiplexing improves the number of operations per second by 4.1 times over spatially multiplexed WDM configurations, positioning it as a strong candidate for next-generation large-scale photonic processors.}


\maketitle

\section{Introduction}\label{sec:intro}

Multiple-input multiple-output (MIMO) links are essential for today’s high-performance communication infrastructure, supporting increased data rates and improved spectral efficiency in wireless and optical networks \cite{winzer2018fiber, wu2024energy}. To process high-speed MIMO signals, analog radio-frequency (RF) or optical signals are often converted into digital signals and handled by electronic digital signal processors (DSP) \cite{cartledge2017digital, morsy2015224}. However, as communication standards evolve toward massive MIMO configurations, the computational throughput and DSP latency become a bottleneck, impeding real-time data processing \cite{hecht2016bandwidth,zhu2022dsp}. These limitations extend to RF signal processing applications, such as interference cancellation \cite{son2020interference}. Electronic processors struggle to meet the evolving demands of these high-throughput applications as they face practical challenges such as bus latency, power leakage, and signal distortion \cite{mack_fifty_2011, lundstrom2022moore}. In contrast, recent developments in photonic processors offer potential solutions to overcome these limitations, particularly in terms of compute density (measured as multiple-accumulate (MAC) operations per second per unit area) and energy efficiency (energy consumed per MAC) \cite{noauthor_femtofarad_nodate, tait_microring_2016}. Photonic processors inherently possess advantages such as low propagation loss, high bandwidth, and resilience to large fan-in and fan-out configurations \cite{noauthor_photonics_nodate, shastri_neuromorphic_2017, noauthor_training_nodate, huang2021silicon}. Moreover, photonic interconnects are free from parasitic effects like resistance or capacitance \cite{nezami2022packaging}. As a result, photonic processors have the potential to achieve energy efficiencies as low as attojoule per MAC, with picosecond latency, and compute densities of petaMAC per second per $\text{mm}^2$ \cite{feldmann_parallel_2021, zhang2023broadband}, surpassing the capabilities of traditional digital processing. Various photonic platforms have been explored, including those based on discrete components, free-space optics, and integrated photonics \cite{wu2023lithography,shen2017deep,teugin2021scalable}. Among these, on-chip photonic processors, which leverage CMOS-compatible fabrication, enable high-density integration and provide low-latency, high-speed, and energy-efficient operations suitable for real-time signal processing \cite{lederman2023real}. Photonic processors utilize multiple degrees of freedom inherent to light, enabling massively parallel and distributed processing through time-division multiplexing (TDM), wavelength-division multiplexing (WDM), and space-division multiplexing (SDM). Recent advancements include TDM-based processors that utilize thin-film lithium niobate modulators \cite{lin202365}, WDM-based processors that employ microring resonators (MRRs) and crossbar arrays in silicon and phase-change photonics \cite{shastri2021photonics, shekhar2024roadmapping}, and SDM-based processors that use silicon Mach-Zehnder interferometer (MZI) meshes \cite{shen2017deep}. In addition, combining mode-division multiplexing (MDM) with WDM has been proposed in fiber-based photonic beamformers to enhance scalability \cite{chang_highly_2013}. However, on-chip photonic processors still face fundamental scalability challenges \cite{tait_neuromorphic_2017, tait_microring_2016} that restrict their use in large-scale applications such as massive MIMO signal processing.

Recent interest in hybrid multiplexing approaches offers promising solutions for performing multiple operations in parallel, leading to higher throughput and enhancing scalability. In particular, hybrid on-chip architectures that combine MDM with WDM offer unparalleled scalability and large bandwidth, significantly expanding the capacity of communication links \cite{luo2014wdm}. In this hybrid approach, n WDM wavelength channels can be reused across m MDM channels, effectively scaling capacity to m × n. However, mode confinement in multimode devices varies significantly among different modes and devices, posing challenges in achieving adequate mode selectivity and mode coupling, which can result in severe intermodal crosstalk and high loss. Similarly, different eigenmodes in multimode waveguides introduce higher dimensional complexity in the waveguide design, particularly in achieving switching functionalities, making controlled mode switching in on-chip interconnect systems \cite{luo2014wdm} highly challenging. 

\begin{figure}[hbt]
    \centering
    \includegraphics[width=1\columnwidth]{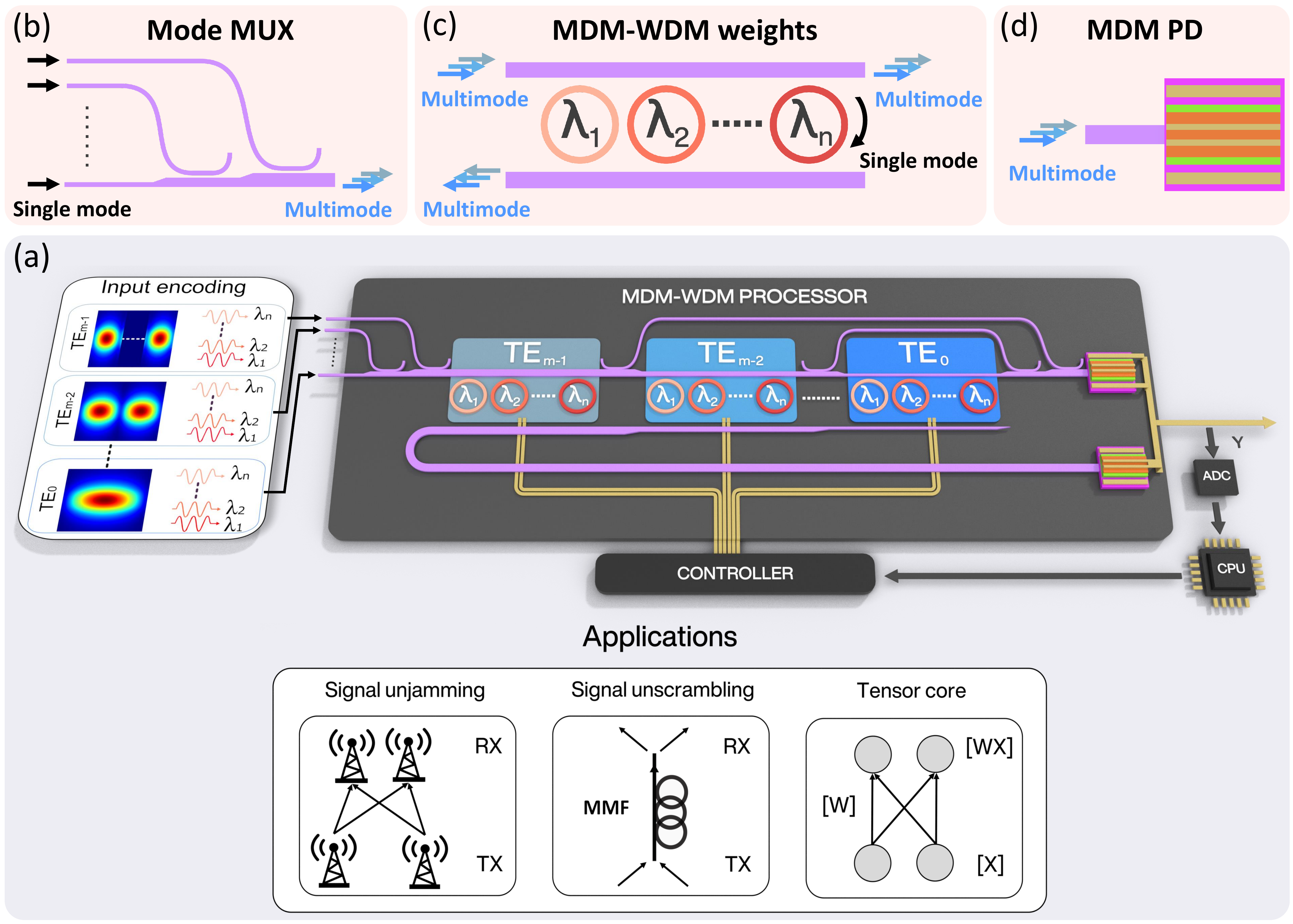}
    \caption{Schematic of the large-scale fully integrated hybrid multimode-multiwavelength photonic processor. (a) The photonic processor is comprised of a mode multiplexer, tunable multimode microring resonators, and multimode balanced photodetectors. Inputs are encoded in both mode and wavelength domains, represented by different colours and field profiles. (b) The mode multiplexer combines single-mode signals into different modes to be input to the processor. (c) The multimode MRR is mode-selective and wavelength-selective by adjusting the bus waveguide width and ring radius, respectively, imprinting the weight matrix on the input signal. The controller drives the MRR heater to tune its resonance wavelength. (d) The integrated multimode photodetectors combine the weighted output signals from the multimode MRR weight bank. The photonic processor is capable of performing a wide range of applications, directly in the mode domain, suitable for applications such as signal unscrambling, signal unjamming, and tensor core processing.}
    \label{fig:mdm_concept}
\end{figure}

Here, we introduce a monolithically integrated, fully functional MDM-WDM-compatible processor that addresses scalability challenges \cite{gordon_design_2017}. The photonic processor, illustrated in Figure \ref{fig:mdm_concept}, comprises an adiabatic mode multiplexer (MUX), a multimode microring resonator array, and multimode balanced photodetectors (PDs). Previous designs for on-chip photonic processors employing MDM, based on all-pass single-mode MRRs and MZIs, have been proposed \cite{yin2023integrated, ling_-chip_2023}; however, these systems offer limited functionalities and multimode compatibility, restricting their versatility to a wide range of applications. Specifically, these architectures neither support negative weighting nor allow multimode signal addition. Additionally, they lack direct mode weighting and rely on demultiplex-weight-multiplex scheme, which is inherently inefficient. Moreover, the MZI-based architecture does not support WDM and uses a passive multimode combiner whose output depends on the relative phases and amplitudes of input signals, creating calibration challenges that may require additional phase shifters. In contrast, our approach integrates multimode WDM-compatible MRR weight banks and multimode germanium photodetectors in a balanced configuration, enabling direct mode weighting and the addition of signals of different wavelengths and modes while facilitating positive and negative weighting. Furthermore, our adiabatic mode multiplexer design alleviates stringent requirements for mode-selective coupling, accommodating large fabrication tolerances. Using this approach, we demonstrate two processors: one with 2 modes and 2 wavelengths, and another with 4 modes and 1 wavelength. Our hybrid multiplexed photonic processor paves the way for high-speed, real-time signal processing applications.

As the channel capacity of transmission links approaches the nonlinear Shannon limit, the growing demand for increased capacity necessitates a new transmission paradigm to manage escalating data traffic. In particular, using MDM with multimode fiber (MMF) in MIMO detection techniques can significantly increase information capacity. However, in conventional MIMO systems, detected subchannels often suffer from mode scrambling due to modal dispersion, scattering, mode mixing, diffraction loss, and crosstalk from neighboring subchannels. This mode scrambling introduces significant decoding complexity, making it challenging to unscramble the transmitted signals \cite{luo_exploiting_2014,ploschner2015seeing,annoni2017unscrambling}. Our fully integrated MDM-WDM-compatible photonic processor overcomes these challenges by enabling arbitrary access to individual modes and wavelengths. This allows for reconfigurable real-time unscrambling of mixed modes in massive MIMO links. As a proof of concept, we demonstrate real-time unscrambling of two modes at a bit rate of 5 Gb/s, with a total processing latency of just 30 ps. We also showcase a scalable application of the reconfigurable photonic processor for RF signal unjamming with a processing bandwidth of 2.5 GHz, thereby showcasing its multifunctional capabilities. To our knowledge, this work represents the first fully integrated monolithic architecture of an MDM-WDM-compatible, all-multimode photonic processor capable of direct negative and positive mode weighting and mode addition via integrated multimode balanced photodetectors. Finally, we show that the hybrid multiplexing processor achieves approximately 4.1 times the performance improvement in the number of operations per second as it scales compared to electrically connecting the outputs of multiple WDM processors, establishing it as a potential candidate for the next-generation large-scale photonic processors.
\section{Results}\label{sec:results}

\subsection{Multimode photonic processor}\label{sec:mux}
\begin{figure}[h!bt]
    \centering
    \includegraphics[width=1\columnwidth]{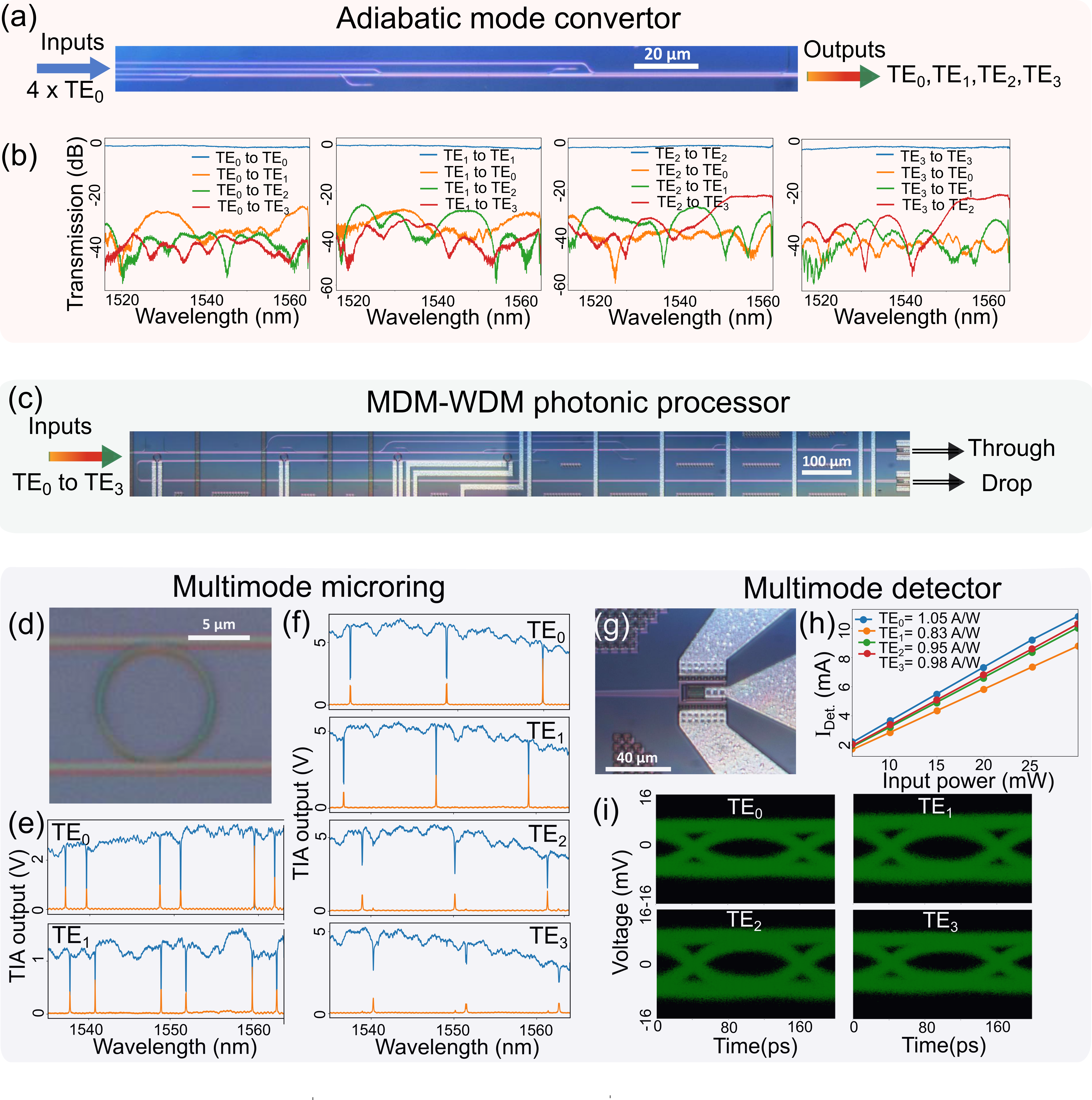}
    \caption{MDM-WDM-compatible Photonic Processor. (a) Image of the four-mode multiplexer. (b) Transmission spectra of the multiplexer measured from back-to-back connected identical multiplexers for different input modes: $\mathrm{TE_0}$, $\mathrm{TE_1}$, $\mathrm{TE_2}$, and $\mathrm{TE_3}$. (c) Image of the photonic processor fabricated on a standard silicon photonic platform. (d) Zoomed image of multimode microring resonator. (e) The through (blue line) and drop (orange line) transmission spectra for $\mathrm{TE_0}$ and $\mathrm{TE_1}$ modes in the 2-mode and 2-wavelength processor. (f) The through (blue line) and drop (orange line) transmission spectra for $\mathrm{TE_0}$, $\mathrm{TE_1}$, $\mathrm{TE_2}$, and $\mathrm{TE_3}$ modes in the 4-mode and 1-wavelength processor. (g) Zoomed image of the designed multimode photodetector. (h) photodetector responsivity for $\mathrm{TE_0}$-$\mathrm{TE_3}$ channels at a reverse bias of 3V. (i) Recorded eye diagrams from the PD for different modes at an operating speed of 10 Gb/s.}
    \label{fig:device}
\end{figure}

The schematic of Figure \ref{fig:mdm_concept} illustrates the MDM-WDM-compatible on-chip photonic processor. The multimode photonic processor comprises three main integral components: an adiabatic mode multiplexer, multimode microring resonators, and multimode photodetectors. The image of the fabricated MDM-WDM-compatible photonic processor, along with the performance of each component, is shown in Figure \ref{fig:device}. In order to realize the hybrid processor, all photonic devices need to be MDM-compatible and simultaneously support multiple WDM channels. Hence, we engineered individual photonic devices in terms of modal loss, crosstalk, and responsivity. The adiabatic directional coupler has a slowly-varying waveguide width to support robust multimode operation, with the index matching point lying in the middle of the coupler (see supplementary section \ref{app:adc_sims}). The integrated adiabatic directional coupler in the processor offers two primary advantages: 1. A guaranteed index-matched coupling point along the coupler to compensate for any fabrication variations \cite{ding_-chip_2013}, and 2. Preventing higher-order modes from coupling back into the single mode in the input arm. It should be noted that since the adiabatic directional coupler works based on index matching to convert a single mode into a higher-order mode in the multimode waveguides, other modes present in the multimode waveguide pass through unaffected. The mode multiplexer/demultiplexer (MUX/deMUX) stage is designed by cascading adiabatic directional couplers. Figure \ref{fig:device}a shows an image of the mode MUX/deMUX. The transmission responses at all output ports are measured individually when light is launched into each of the input ports to estimate the insertion loss and crosstalk of each of the mode MUX/deMUX ports. The transmission spectra (Figure \ref{fig:device}b) show broadband performance over more than 30 nm wavelength range with an insertion loss of 1.9, 1.5, 1.5, 3 and maximum crosstalk of 33, 28, 28, 26 at 1550 nm for TE$_{0}$, TE$_{1}$, TE$_{2}$ and TE$_{3}$, respectively. The parameters of the designed adiabatic directional coupler are given in Table \ref{Tb:adc}. As a proof of concept, we designed a four-mode MUX/deMUX, however, it can accommodate more than 10 modes with comparable performance \cite{dai_10-channel_2018}. Furthermore, by utilizing photonic inverse design algorithms \cite{christiansen_inverse_2021}, it is possible to realize an efficient TE$_{0}$ to TE$_{1}$ mode multiplexer with a more compact footprint \cite{add_yang_multi-dimensional_2022}. 

The weight bank in the photonic processor is composed of an array of multimode microring resonators. Unlike conventional single-mode microring resonator arrays, the engineered multimode microring resonators (Figure \ref{fig:device}d) have both mode-selective (TE$_{0}$ to TE$_{3}$) and wavelength-selective features. Each ring is designed with different bus waveguide width and ring radius, which determine the appropriate mode and resonance wavelength, respectively \cite{noauthor_two-mode_nodate}. The rings are precisely engineered using mode index curves to ensure that only the correct mode couples strongly into the ring. A detailed discussion of the ring design and parameters is provided in the supplementary section \ref{app:mm_ring}. The through and drop responses were measured and are shown in Figure \ref{fig:device}e for a 2-mode and 2-wavelength weight bank and in Figure \ref{fig:device}f for a 4-mode and 1-wavelength weight bank. Evidently, each ring is interacting with the correct mode and correct wavelength. Finally, to realize a fully integrated photonic processor, we have integrated a balanced multimode PIN germanium-on-silicon vertical photodetectors (BPD) at the end of the through and drop multimode waveguides. The image of the multimode photodetector is shown in Figure \ref{fig:device}g. For efficient multimode operation, the BPD is carefully designed with two off-centred metal contacts on top of the N-doped section instead of one centred contact (see supplementary section \ref{sec:detectors}). This results in less attenuation near the peaks of modal power profiles, thereby improving detector responsivity \cite{fard_responsivity_2016} for all modes. The designed multimode PD has an active area of 10.6 $\mu$m x 27.2 $\mu$m with responsivities ranging from 0.83 to 1.05 A/W for TE$_{0}$ to TE$_{3}$ modes (Figure \ref{fig:device}h). In addition, the recorded eye diagrams (Figure \ref{fig:device}i) validate the high-speed functionality of the PD across all modes. PIN germanium-on-silicon vertical photodetectors can achieve operation bandwidths exceeding 50 GHz through careful dimension optimization and the use of inductive peaking \cite{fard_responsivity_2016, zhu_high_2018}. The fully integrated MDM-WDM-compatible on-chip photonic processor is demonstrated for 2-mode/2-wavelength channels, and 4-mode/1-wavelength channels (Figure \ref{fig:device}c), incorporating mode MUX, MDM microring resonators, and multimode photodetectors suitable for a wide range of reconfigurable high-speed real-time signal processing applications.

\subsection{Optical signal unscrambling}\label{sec:unscramble}
 \begin{figure}[hbt]
    \centering
    \includegraphics[width=1\columnwidth]{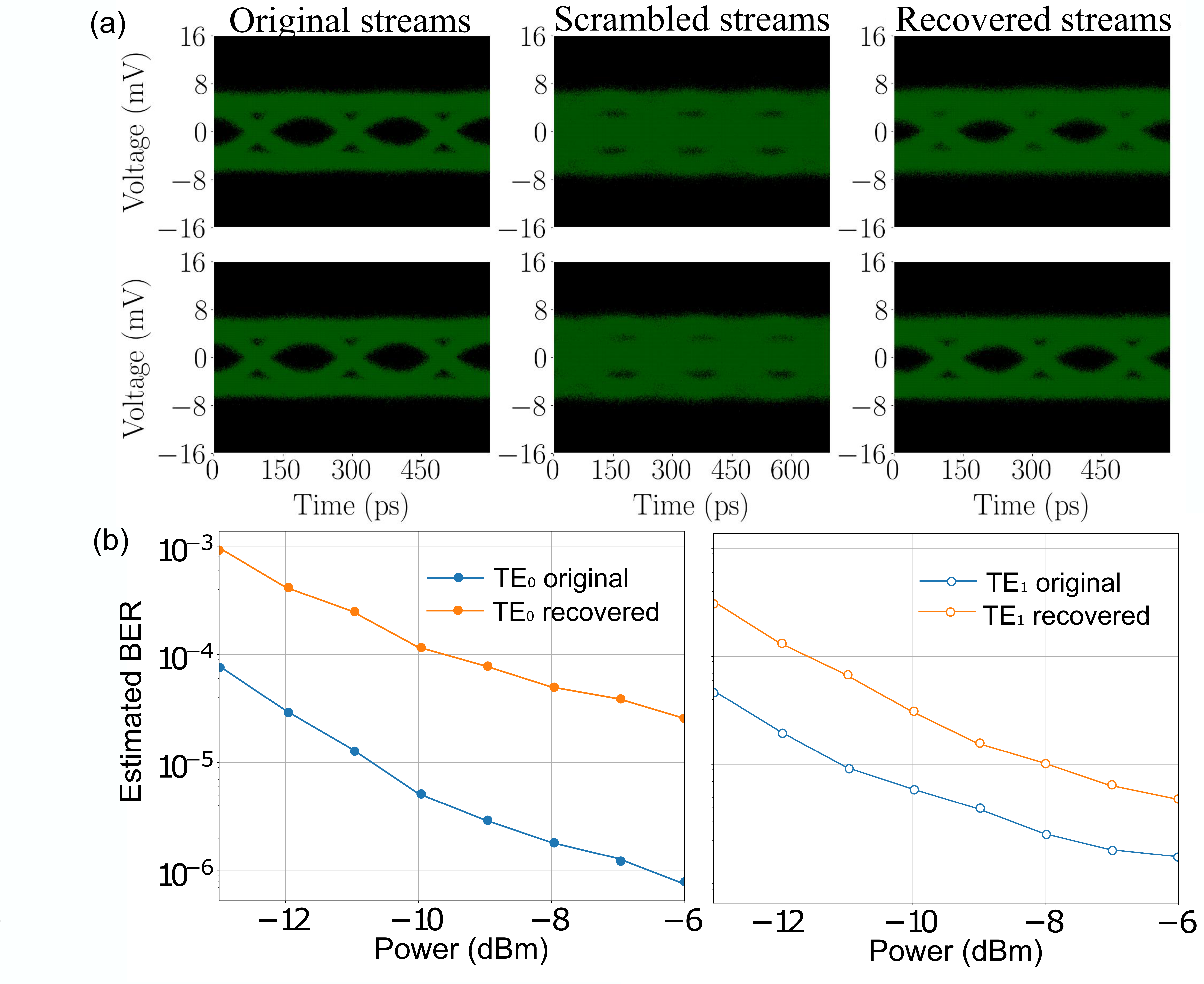}
    \caption{Experimental demonstration of unscrambling optical data streams: Two emulated scrambled signals are encoded on two mode channels ($\mathrm{TE_0}$ and $\mathrm{TE_1}$) and sent to the photonic processor at an operating speed of 5 Gb/s. (a) Recorded eye diagrams of the original NRZ streams (S$_{\mathrm{TE_0}}$ and S$_{\mathrm{TE_1}}$), the scrambled NRZ streams (X$_{\mathrm{TE_0}}$ and X$_{\mathrm{TE_1}}$), and the unscrambled NRZ streams ($\mathrm{\hat{S}}_{\mathrm{TE_0}}$ and $\mathrm{\hat{S}}_{\mathrm{TE_1}}$) using the photonic processor. The first row shows the unscrambling of the $\mathrm{TE_0}$ mode, and the second row shows the scrambling of the $\mathrm{TE_1}$ mode, respectively. (b) Estimated BER versus input power for the recovered $\mathrm{TE_0}$ and $\mathrm{TE_1}$ streams compared to the original signals. The photodetectors are reverse biased at 3V using a bias tee, and the signal outputs are recorded using a high-speed scope.}
    \label{fig:unscramble}
\end{figure}

Conventional MIMO-based communication links demand a high level of digital signal processing (DSP) complexity to unscramble the transmitted signals, leading to increased latency, energy consumption, and device bulkiness \cite{chen2012flexdet,barbosa_scalable_2022,arik2014mimo,cristiani_roadmap_2022}. However, our designed photonic chip can directly unscramble outputs in real time for high-speed MIMO transmission links. Unlike other processors, our multimode photonic processor is compatible with applications requiring direct processing of inputs in the mode domain. The details of the experimental setup are in the supplementary section \ref{app:setup}. As a proof of concept, we designed the experiment for real-time unscrambling of two modes, TE$_{0}$ and TE$_{1}$, operating at 5 Gb/s. We generated two RF signals from an arbitrary waveform generator (AWG) to emulate the scrambling of optical signals equivalent to the multimode fiber transmission link. This way, we can achieve a controlled generation of scrambled output signals by applying predefined scrambled modulation to the original signal. In the experiment, we scrambled two raised cosine non-return-to-zero (NRZ) pseudorandom binary sequence (PRBS) signals of length 2$^{13}$-1 then encoded the result on two mode signals (TE$_0$ and TE$_1$) of the same wavelength using Mach-Zehnder modulators (MZM). Subsequently, the scrambled signals are sent to the 2-mode and 2-wavelength processor multimode photonic processor for real-time unscrambling. Typically, the scrambling is mode-dependent and Baud rate-dependent, which has a substantial impact on system performance and bit error rate (BER). By sending pilot signals \cite{zhao_digital_2015}, which are a set of signals known a priori by the receiver, we are able to estimate the scrambling matrix (See Methods). The unscrambling is efficiently performed via a matrix-vector multiplication (MVM) where the inverse of the scrambling matrix is applied to the mode channels using the mode-selective microring resonator weight bank, then summing them at the detectors.

Figure \ref{fig:unscramble} shows eye diagrams of the original, scrambled, and recovered signals. The scrambled signal, operating at 5 Gb/s, displays a closed eye. Upon applying the inverse of the scrambling matrix using the thermally-tuned microring resonator weights, our photonic processor successfully unscrambles the transmitted signal, displaying a visible eye-opening as evident from Figure \ref{fig:unscramble}a with BER of $2.6 \times 10^{-5}$ and $5.7 \times 10^{-6}$ for TE$_0$ and TE$_1$ channels, respectively, at the input power value of -6 dBm. Figure \ref{fig:unscramble}b illustrates the estimated BER with input power values for the two different mode signals. The TE$_0$ and TE$_1$ modes exhibit a power penalty of approximately 6 dB and 3 dB, respectively, at the input power value of -6 dBm. This power penalty primarily arises from modal crosstalk and microring weights bit precision, which can be further reduced by optimizing device performance. The photonic device demonstrates no resistive or capacitive effects, making it highly energy-efficient and low-latency. The system has an estimated total processing latency of 30 ps, making it well-suited for high-speed real-time data transmission and communication applications.

\subsection{RF signal unjamming}\label{sec:unjam}
 \begin{figure}[hbt]
    \centering
    \includegraphics[width=1\columnwidth]{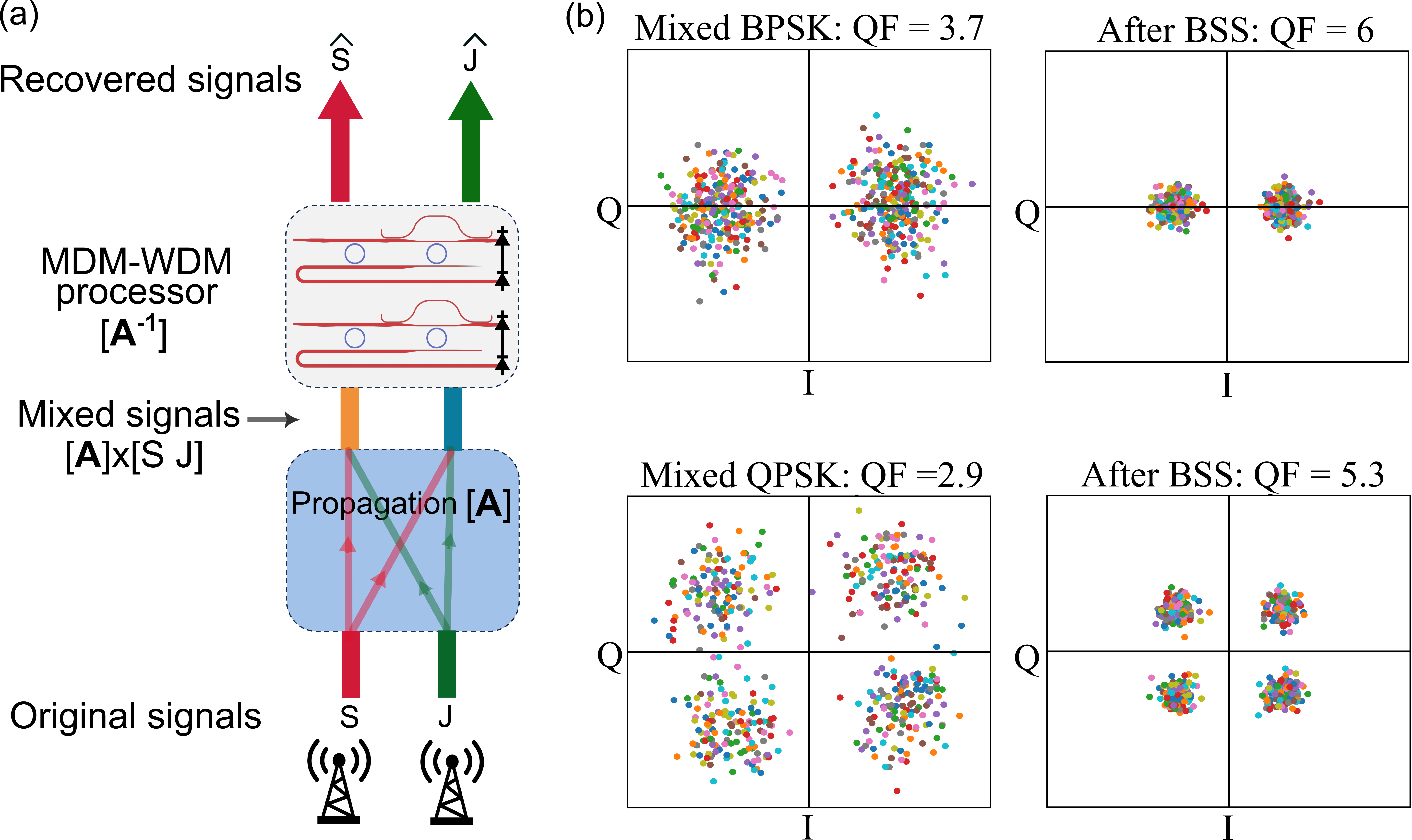}
    \caption{Experimental demonstration of signal unjamming: (a) Schematic of signal unjamming through the photonic processor. S represents the phase-encoded original BPSK and QPSK signals, while J represents the jamming signal. Both are emulated using intensity modulators. The signals are mixed via the propagation matrix \textbf{A}. The input signals are fed into $\mathrm{TE_{0}}$ and $\mathrm{TE_{1}}$ channels of the multimode processor. Upon implementing the photonic weight matrix \textbf{A$^{-1}$} via the MRRs of the processor, the original signal $\mathrm{\hat{S}}$ is separated from the jammer $\mathrm{\hat{J}}$. (b) The constellation diagram shows the jammed and unjammed BPSK and QPSK signals respectively, along with their Q-factors (QF).}
    \label{fig:unjam}
\end{figure}

Finally, in order to demonstrate the versatility of our system, we performed RF signal unjamming using blind source separation (BSS). Typically, noisy RF interference is one of the most significant issues in modern communication systems, especially with the increasing number of antennas and channel bandwidth \cite{zhang2024system}. Even though digital post-processing methods have proven efficient for interference cancellation in low-speed, narrow-band applications, they still face significant latency and energy efficiency challenges at GHz operation. Conversely, incorporating the combination of a multimode microring weight bank and integrated photodetectors provides the direct advantage of on-chip, scalable, and low-latency photonic BSS operating at a high processing bandwidth. The block diagram of RF signal unjamming is shown in Figure \ref{fig:unjam}a. We emulate the mixing of the signal and jammer signal from the MIMO antennas in the RF domain using two AWG channels then feed them to two MZMs operating at the same wavelength. The two resultant optical signals are encoded on TE$_0$ and TE$_1$ modes in our 2-mode and 2-wavelength processor. The phase-shift keying (PSK) signal is transmitted at a bit rate of 2.565 GHz on a carrier of the same frequency. On the other hand, the broadband jamming signal is generated as random values from a continuous uniform distribution. The mixed signals can be represented by \textbf{X} = \textbf{AS}, where \textbf{A} is the mixing matrix, and \textbf{X} and \textbf{S} are the mixed and original signal vectors, respectively. Following the relation \textbf{S} = \textbf{A$^{-1}$X}, which is applying the inverse matrix to the mixed signals, the original unmixed signals vector can be retrieved at the receiver side. We implemented BSS using our integrated multimode photonic processor as a powerful technique for cancelling noisy interference with minimal information about the communication link \cite{han2015large}.

In order to experimentally recover the original PSK signal, we first directly recorded the received data in each mode channel at the multimode photodetectors. Following that, we performed independent component analysis using the FastICA \cite{hyvarinen_fast_1999,hyvarinen_independent_2000} Python package on the recorded signal to estimate the mixing matrix (see Methods). Next, we used tunable heaters on each microring resonator weight bank to implement the inverse matrix. Figure \ref{fig:unjam}b displays the constellation diagrams of the jammed and unjammed binary PSK (BPSK) and quadrature PSK (QPSK) signals, along with their Q-factors. As shown in the figure, after unjamming, the noise in the constellation diagrams is noticeably reduced, leading to significant improvements in the Q-factor values from 3.7 to 6.0 and from 2.9 to 5.3 for BPSK and QPSK signals, respectively. The BSS algorithm is implemented based on kurtosis analysis, making it suitable for both coherent and incoherent communication links regardless of carrier frequency, signal format, and channel conditions. Our demonstration showcases the recovery of the original signal from the jammer signal in real-time over broadband coverage, enhancing efficient data transmission in MIMO RF communication links.
\section{Discussion}\label{sec:disc}
\begin{figure}[hbt]
    \centering
    \includegraphics[width=0.8\columnwidth]{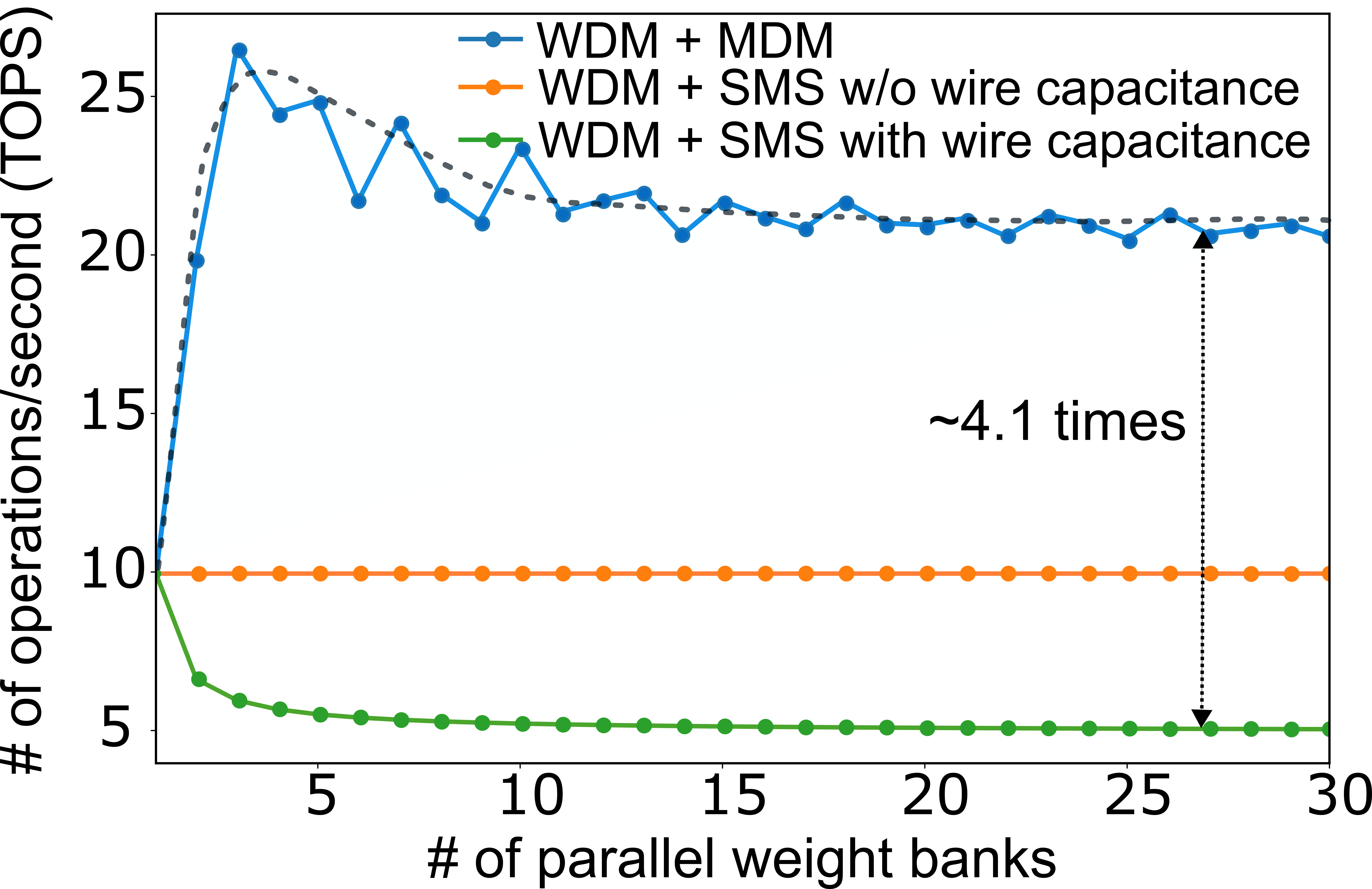}
    \caption{Comparison of the number of tera operations per second (TOPS) with up-scaling between WDM-SMS and WDM-MDM architectures. In the WDM-SMS approach, the processing speed decreases with system scaling due to increased wiring capacitance. Assuming negligible wiring capacitance, TOPS remains constant with up-scaling. However, in the MDM-WDM architecture, processing speed increases with scaling due to the larger number of supported modes, despite the speed reduction caused by increased junction capacitance. The dotted line is a guide to the eye.}
    \label{fig:scalability}
\end{figure}
Large-scale photonic processors are needed for many applications such as massive MIMO systems \cite{noauthor_mimo_nodate-1} and large-scale MVM in neural networks. To scale photonic processors beyond WDM limits that are imposed by WDM filters, we can combine multiple instances of WDM processor with spatial multiplexing scheme (SMS) or MDM as we suggest in this work. Spatial multiplexing (supplementary Figure \ref{fig:scale-sms}) entails connecting the PDs of parallel WDM processors in order to combine the output currents. However, this up-scaling comes at the cost of reduced operation speed of the PDs due to increasing the node capacitance at the output. This node capacitance increase is due to the accumulation of the PD capacitances in addition to wiring capacitance. This can be described by the following equation:
\begin{equation}
\label{eq:ops_sms}
    \mathrm{{OPS}_{SMS}} = M(2N-1)F\frac{C_{PD}}{MC_{PD}+(M-1)C_W},
\end{equation}
where $\mathrm{{OPS}_{SMS}}$ is the number of operations per second when employing SMS to combine WDM processors, $M$ is the number of combined WDM processor instances, $N$ is the number of wavelengths employed in the WDM processor, $C_{PD}$ is the junction capacitance of a reference PD \cite{fard_responsivity_2016}, $F$ is its operation speed and $C_W$ is the estimated wiring capacitance required to connect two PDs. The parameter values are listed in supplementary section \ref{app:sms}.

The alternative that we suggest in this work is to do this combination in the optical domain by combining WDM processor instances using MDM (supplementary Figure \ref{fig:scale-mdm}). To increase the number of supported modes in a waveguide, we need to increase the waveguide width and we consequently, increase the PD width to maintain maximum responsivity. This PD width increase comes at the expense of reduced speed due to increased junction capacitance, however, the increase in the number of supported modes more than compensates for it. In addition, there is no wiring capacitance involved in this case. This can be described using the following equation:
\begin{equation}
\label{eq:ops_mdm}
    \mathrm{{OPS}_{MDM}} = M(2N-1)F\frac{C_{PD}^{SM}}{C_{PD}^{MM}(M)},
\end{equation}
where $\mathrm{{OPS}_{MDM}}$ is the number of operations per second when employing MDM to combine WDM processors, $M$ is the number of combined WDM processor instances and also represents the number of employed modes, $C_{PD}^{SM}$ is the single-mode PD capacitance and $C_{PD}^{MM}(M)$ is the multimode PD capacitance as a function of $M$. The number of supported modes at each waveguide width, and hence PD capacitance, was computed using the Lumerical Mode software package \cite{lumerical}. The number of operations per second (OPS) versus the number of combined WDM processor instances using SMS and MDM techniques is reported in Figure \ref{fig:scalability}. As demonstrated, the OPS for SMS decreases at the expense of up-scaling. At best, it remains constant with up-scaling if negligible wiring capacitance is assumed, which is usually not the case. On the contrary, not only does the MDM approach conserve the OPS while up-scaling, but it also increases it, which highlights the significance of up-scaling using MDM.
\section{Conclusion}\label{sec:conc}
Our demonstration is the first implementation of a monolithically integrated MDM-WDM-compatible processor comprising adiabatic mode multiplexers, multimode microring resonator weight bank, and multimode balanced photodetectors. By integrating mode-division multiplexing and wavelength-division multiplexing, we have advanced the scalability and throughput of photonic processors. We demonstrated the versatility of our MDM-WDM-compatible processor for two real-time applications: optical MIMO signal unscrambling and RF signal unjamming. The on-chip system has a total processing latency of 30 ps. Furthermore, our analysis shows that the compact and reconfigurable photonic processor can support a very large number of modes and wavelengths, enabling large-scale real-time signal processing applications. Moreover, unlike spatial multiplexing schemes, MDM enables up-scaling while enhancing the system throughput, thereby opening avenues for large-scale high-speed real-time signal processing applications.

\backmatter
\newpage
\section*{Methods}

\setcounter{figure}{0}
\setcounter{table}{0}

\section*{Pilot signaling}\label{app:pilots}
Mixed signals can be recovered by multiplying them with the inverse of the mixing matrix. One way to estimate the elements of the unmixing matrix is to use a set of known a priori signals at the receiver, i.e., pilot signals. The original signals can be recovered from the mixed signals using the equation:

\begin{equation}
    \mathbf{S} = \mathbf{A}^{-1} \mathbf{X} = \mathbf{W} \mathbf{X},
\end{equation}
where $\mathbf{S}$ is the matrix containing the vectors of the original signals $\mathbf{S_1}$ and $\mathbf{S_2}$, $\mathbf{A}$ is the mixing matrix, $\mathbf{W}$ is the unmixing matrix, and $\mathbf{X}$ is the matrix containing the vectors of the mixed signals $\mathbf{X_1}$ and $\mathbf{X_2}$. By expanding the unmixing matrix:
\begin{subequations}
\begin{align}
    \quad &     \begin{bmatrix}
                \mathbf{S_1} \\ \mathbf{S_2}
                \end{bmatrix}
                =
                \begin{bmatrix}
                w_{11} & w_{12} \\
                w_{21} & w_{22}
                \end{bmatrix}
                \begin{bmatrix}
                \mathbf{X_1} \\ \mathbf{X_2}
                \end{bmatrix}, \\
    \quad & \mathbf{S_1}^{p_1} = w_{11}\mathbf{X_1}^{p_1} + w_{12}\mathbf{X_2}^{p_1}, \label{eq:pilot1}\\
    \quad & \mathbf{S_2}^{p_1} = w_{21}\mathbf{X_1}^{p_1} + w_{22}\mathbf{X_2}^{p_1}, \\
    \quad & \mathbf{S_1}^{p_2} = w_{11}\mathbf{X_1}^{p_2} + w_{12}\mathbf{X_2}^{p_2}, \\
    \quad & \mathbf{S_2}^{p_2} = w_{21}\mathbf{X_1}^{p_2} + w_{22}\mathbf{X_2}^{p_2} \label{eq:pilot4}
\end{align}
\end{subequations}
where $w_{11}, w_{12}, w_{21}$, and $w_{22}$ are the elements of the unmixing matrix, $\mathbf{X_1}^{p_1}$ and $\mathbf{X_2}^{p_1}$ are the received signals when $\mathbf{S_1}^{p_1}$ and $\mathbf{S_2}^{p_1}$ pilot signals are transmitted, and $\mathbf{X_1}^{p_2}$ and $\mathbf{X_2}^{p_2}$ are the received signals when $\mathbf{S_1}^{p_2}$ and $\mathbf{S_2}^{p_2}$ pilot signals are transmitted. By solving the equations \ref{eq:pilot1}-\ref{eq:pilot4} simultaneously, we can estimate all the elements of the unmixing elements.

\section*{Blind source separation}\label{app:bss}
Blind source separation (BSS) is a powerful technique used to unmix any uncorrelated mixed signals without prior knowledge about the signals. According to the central limit theorem, the Gaussianity of the mixed signals is always higher than that of the original signals. Therefore, the BSS algorithm searches for the unmixing matrix that maximizes the non-Gaussianity of the unmixed signals. Kurtosis is the metric used to quantify non-Gaussianity, and it is given by:
\begin{equation}
    K[\mathbf{S_i}] = \frac{E[(\mathbf{S_i}-\mu_i)^4]}{\sigma_i^4} - 3,
\end{equation}
where $K$ is the kurtosis of the signal $\mathbf{S_i}$, and $\mu_i$ and $\sigma_i$ are its mean and standard deviation, respectively. The kurtosis of a Gaussian distribution is zero. For BSS, it is often desirable to whiten the received signals by uncorrelating the signals, zeroing their mean, and normalizing their variance. This whitening process ensures that the vectors of the unmixing matrix are orthogonal.

Starting from the received signals $\mathbf{X_1}$ and $\mathbf{X_2}$:
\begin{equation}
    \mathbf{X_c} = \mathbf{X} - \begin{bmatrix} E[\mathbf{X_1}] \\ E[\mathbf{X_2}] \end{bmatrix},
\end{equation}
where $\mathbf{X}$ is a matrix with rows representing the received signals, $\mathbf{X_1}$ and $\mathbf{X_2}$, while $\mathbf{X_c}$ is a matrix containing the corresponding received signals that have been centered (i.e., with zero mean). The centered signals are then uncorrelated and their variance is normalized as follows:
\begin{equation}
    \mathbf{X_w} = \mathbf{V} \mathbf{D}^{\frac{-1}{2}} \mathbf{V}^T \mathbf{X_c},
\end{equation}
where $\mathbf{X_w}$ denotes the matrix comprising rows of the whitened received signals, $\mathbf{D}$ is the diagonal matrix containing the eigenvalues of the covariance matrix of the centered signals, and $\mathbf{V}$ is the matrix of eigenvectors of the covariance matrix of the centered signals. Since the original signals $\mathbf{S_1}$ and $\mathbf{S_2}$ are uncorrelated, we can deduce that the vectors of the unmixing matrix are orthogonal:
\begin{equation}
\label{eq:orthog}
    Cov[\mathbf{S_1} \mathbf{S_2}^T] = E[\mathbf{S_1} \mathbf{S_2}^T] = 0 = E[(\mathbf{w_1} \mathbf{X_w})(\mathbf{X_w}^T\mathbf{w_2}^T)] = \mathbf{w_1}E[\mathbf{X_w} \mathbf{X_w}^T]\mathbf{w_2}^T = \mathbf{w_1} \mathbf{w_2}^T,
\end{equation}
where $\mathbf{w_1}$ and $\mathbf{w_2}$ are the rows of the unmixing matrix $\mathbf{W}$ that are used to recover $\mathbf{S_1}$ and $\mathbf{S_2}$ from the whitened received signals, respectively. It is noteworthy that we assume the mean of the original signals to be zero, so the recovered signals will have zero mean.

Finally, to recover the mixed signals, we begin with random vectors $\mathbf{w_1}$ and $\mathbf{w_2}$, then iteratively optimize them using gradient descent until convergence to a maximum kurtosis as follows:
\begin{subequations}
\begin{align}
    K[\mathbf{S_1}] &= K[\mathbf{w_1} \mathbf{X_w}] = \frac{E[(\mathbf{w_1} \mathbf{X_w})^4]}{\sigma_1^4} - 3, \\
    \intertext{the gradient of this kurtosis is obtained as:}
    \frac{\partial K}{\partial \mathbf{w_1}} &= \frac{3E[(\mathbf{w_1} \mathbf{X_w})^3 \mathbf{X_w}]}{\sigma_1^4}, \\
    \intertext{and $\mathbf{w_1}$ update rule is formulated as:}
    \mathbf{w_1}^{i+1} &= \mathbf{w_1}^{i} - 3 \eta \frac{E[(\mathbf{w_1} \mathbf{X_w})^3 \mathbf{X_w}]}{\sigma_1^4},
\end{align}
\end{subequations}
where $i$ represents the index of the optimization iteration, $\eta$ is the learning rate, and $\sigma_1$ is the standard deviation of $\mathbf{w_1} \mathbf{X_w}$.

Since $\mathbf{w_1}$ and $\mathbf{w_2}$ are orthogonal due to whitening the received signals (equation \ref{eq:orthog}), we can find $\mathbf{w_2}$ using the relation:
\begin{equation}
    \mathbf{w_2} = \mathbf{w_2} - (\mathbf{w_1} \mathbf{w_2}^T)\mathbf{w_1}.
\end{equation}
The original signals are then recovered using the optimized weight vectors as follows:
\begin{subequations}
\begin{align}
    \mathbf{S_1} &= \mathbf{w_1} \mathbf{X_w}, \\
    \mathbf{S_2} &= \mathbf{w_2} \mathbf{X_w}.
\end{align}
\end{subequations}

\bmhead{Supplementary information}
Supplementary information is available for this work.

\bmhead{Acknowledgments}
We acknowledge the Natural Sciences and Engineering Research Council of Canada (NSERC) for their support and CMC Microsystems for providing access to fabrication foundries and design tools.

\section*{Declarations}
\begin{itemize}
\item Funding - We acknowledge support from the Natural Sciences and Engineering Research Council of Canada (NSERC).
\item Conflict of interest/Competing interests - B.J.S cofounded Milkshake Technology Inc.
\item Ethics approval - Not applicable
\item Consent to participate - Not applicable
\item Consent for publication - Not applicable
\item Availability of data and materials - Available on reasonable request
\item Code availability - Available upon reasonable request
\item Authors' contributions - A.K, C.H, A.N.T, and B.J.S conceived the idea. A.K performed the simulations, the layout of the chip, and carried out the experiments. A.N.T reviewed the layout. A.K and A.A wrote the manuscript. All co-authors analyzed and discussed the data and contributed to revising the manuscript. A.N.T and B.J.S supervised this work.
\end{itemize}

\noindent 

\bibliography{sn-bibliography}

\appendix

\newpage
\setcounter{page}{1} 
\section*{Supplementary information}
\makeatletter
\renewcommand \thesection{S\@arabic\c@section}
\renewcommand\thetable{S\@arabic\c@table}
\renewcommand \thefigure{S\@arabic\c@figure}
\renewcommand \theequation{S\@arabic\c@equation}
\makeatother

\setcounter{equation}{0}
\setcounter{figure}{0}
\setcounter{table}{0}

\section{Adiabatic directional coupler optimization}\label{app:adc_sims}
\begin{figure}[hbt]
    \centering
    \includegraphics[width=1.0\columnwidth, height=4cm]{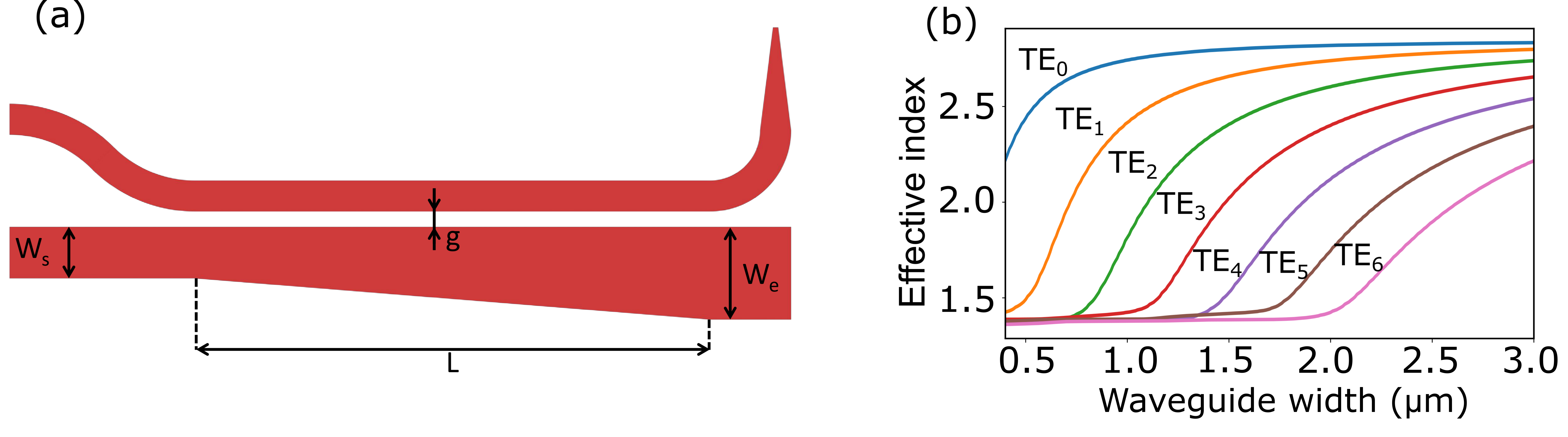}
    \caption{(a) Adiabatic directional coupler structure. The multimode waveguide is tapered from a start width ($W_s$) to an end width ($W_e$), where L and g are the coupler length and gap respectively. This ensures that the mode matching point lies in the coupler even if fabrication variations exist. (b) TE mode effective indices in an SOI strip waveguide of 220 nm depth.}
    \label{fig:adc}
\end{figure}

\begin{figure}[hbt] 
\subfloat[\vspace{0pt}]{\includegraphics[width=0.3\columnwidth]{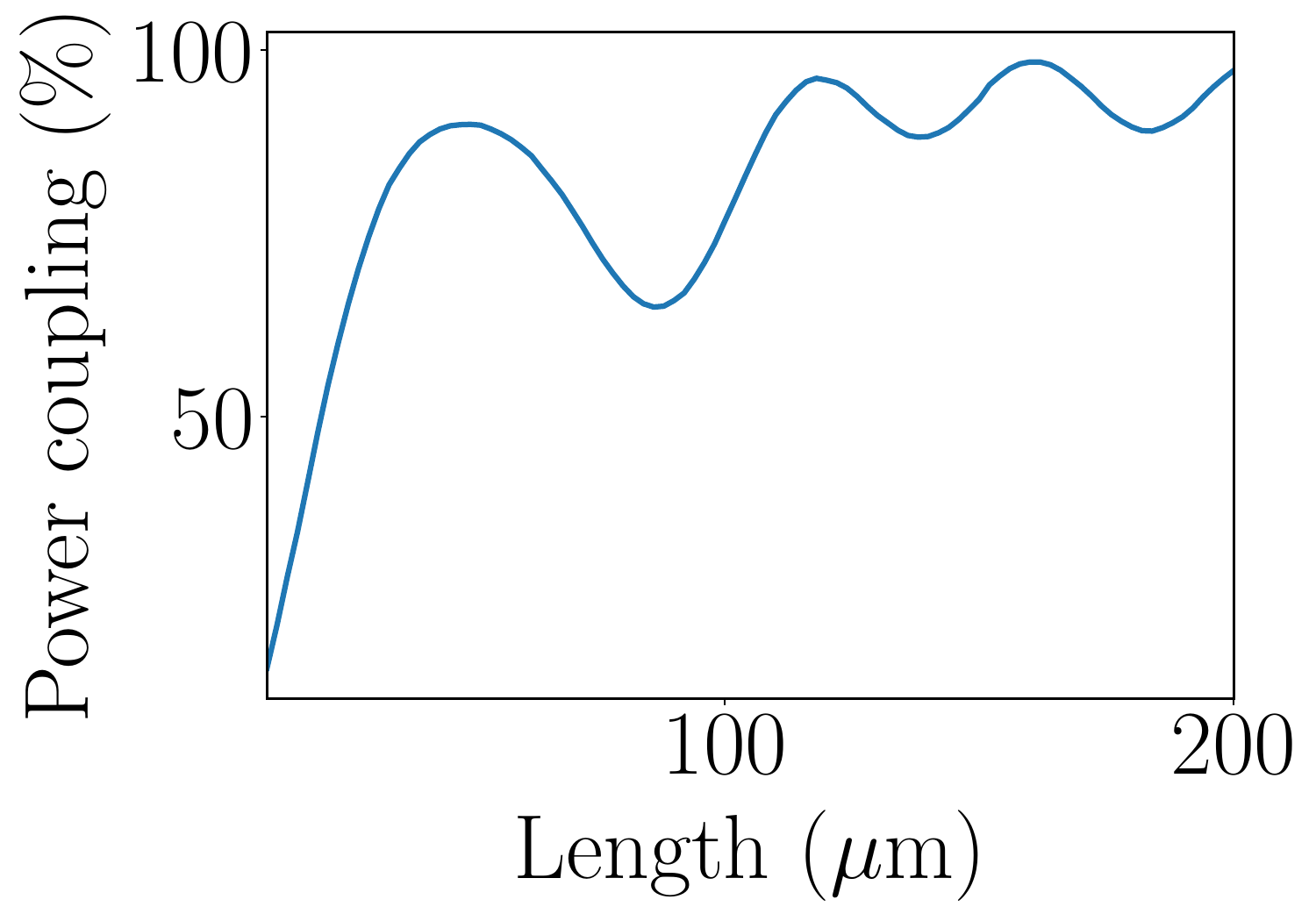}}\quad
\subfloat[\vspace{0pt}]{\includegraphics[width=0.3\columnwidth]{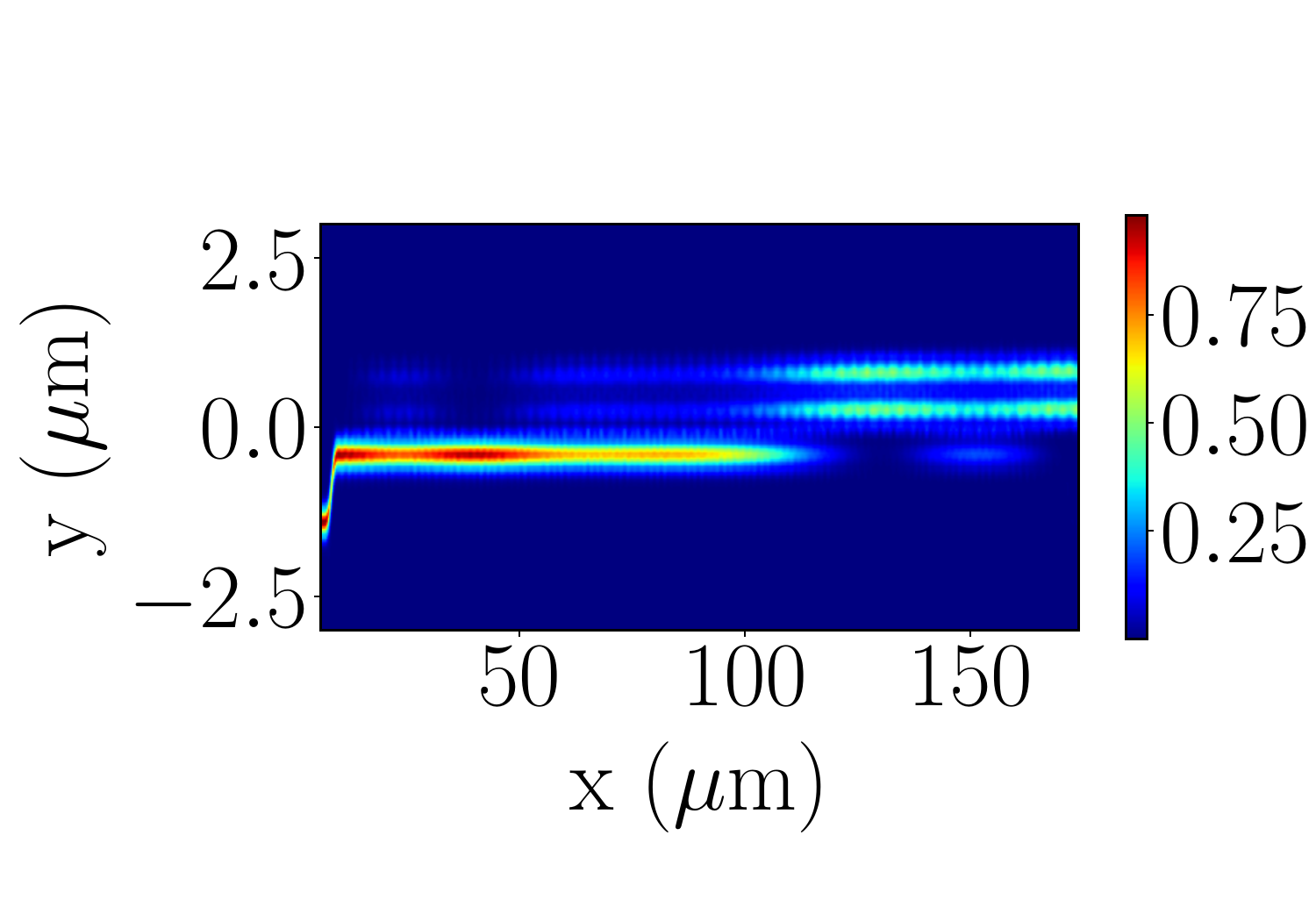}}\quad
\subfloat[\vspace{0pt}]{\includegraphics[width=0.3\columnwidth]{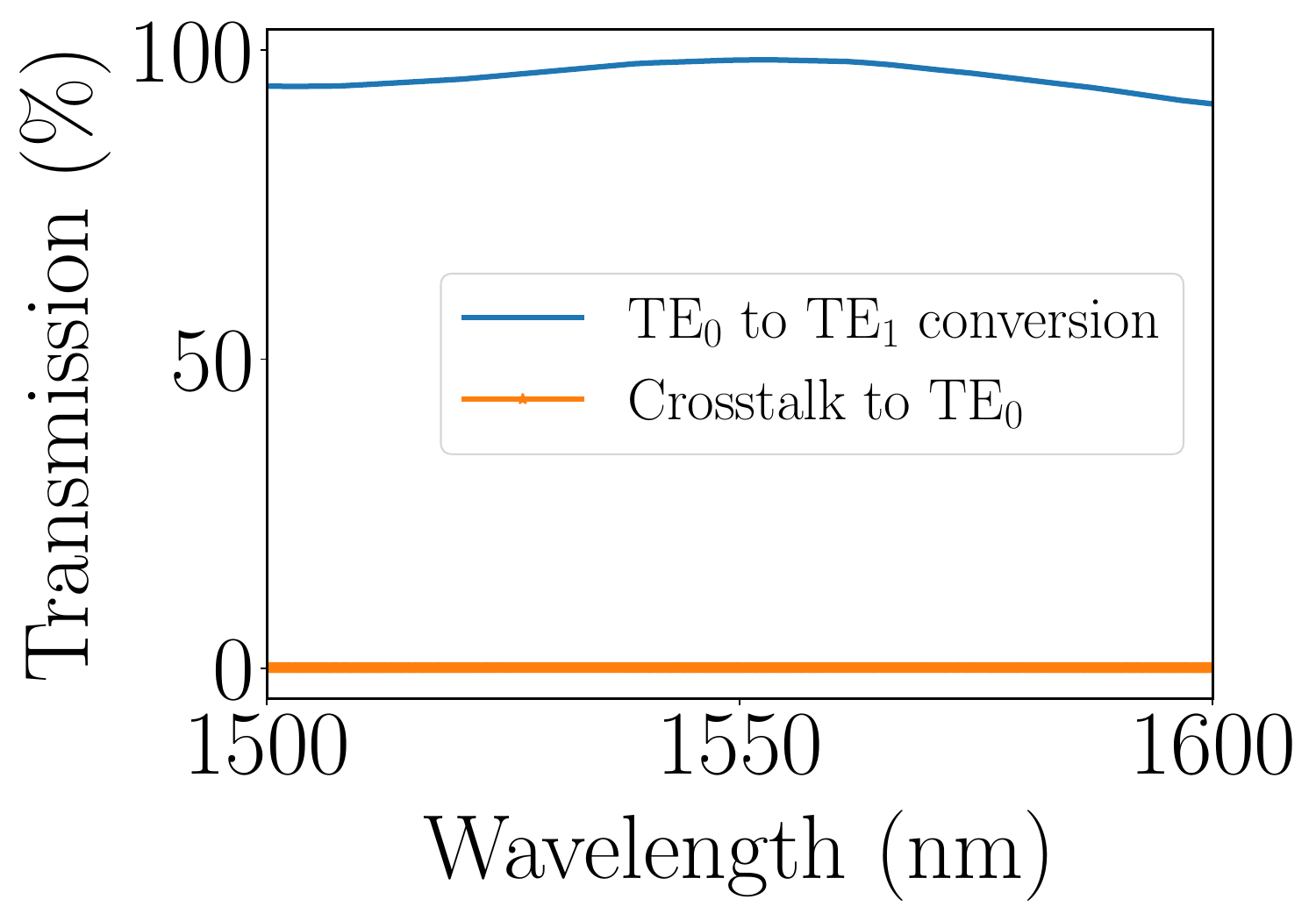}}\quad
\subfloat[\vspace{0pt}]{\includegraphics[width=0.3\columnwidth]{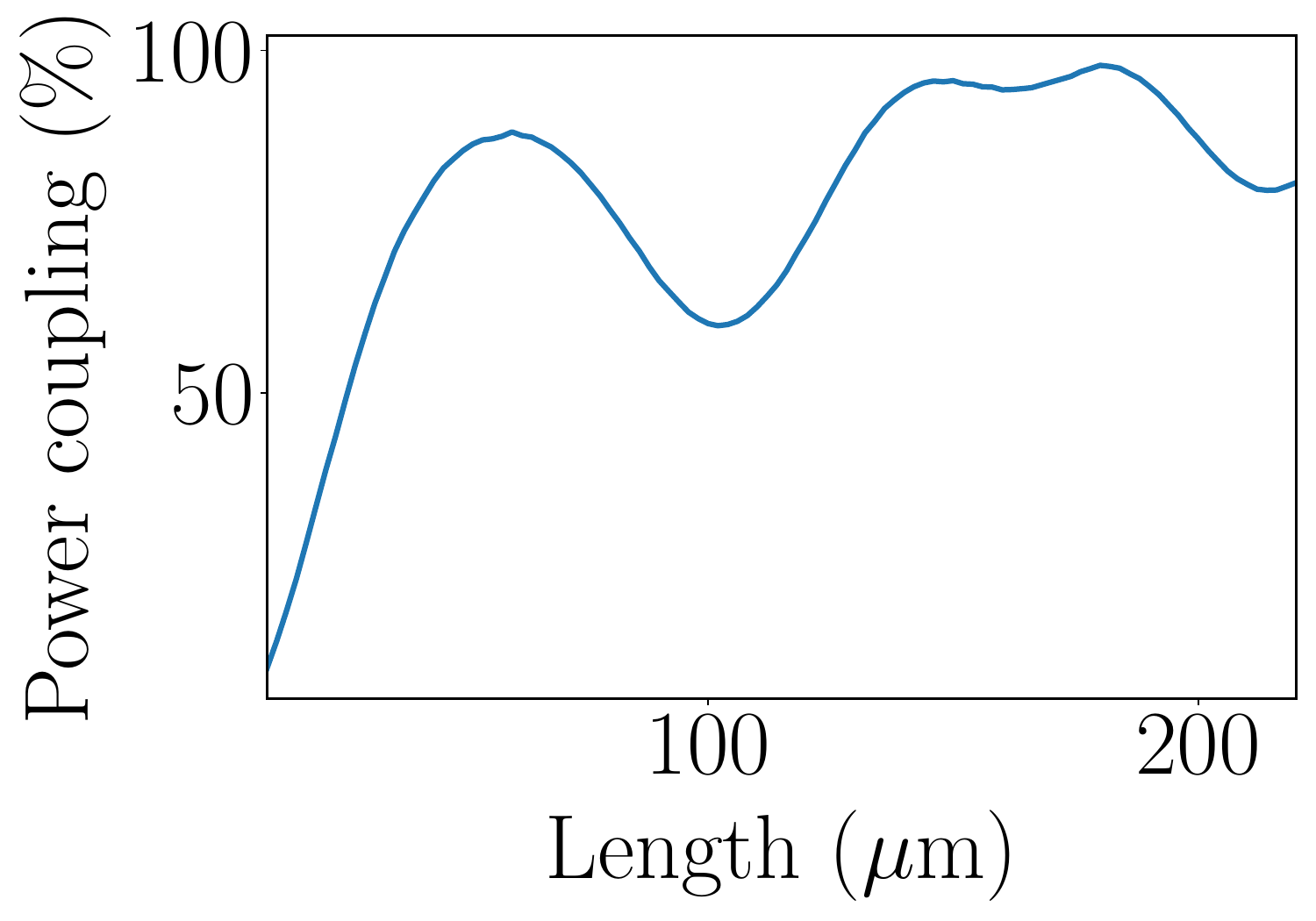}}\quad
\subfloat[\vspace{0pt}]{\includegraphics[width=0.3\columnwidth]{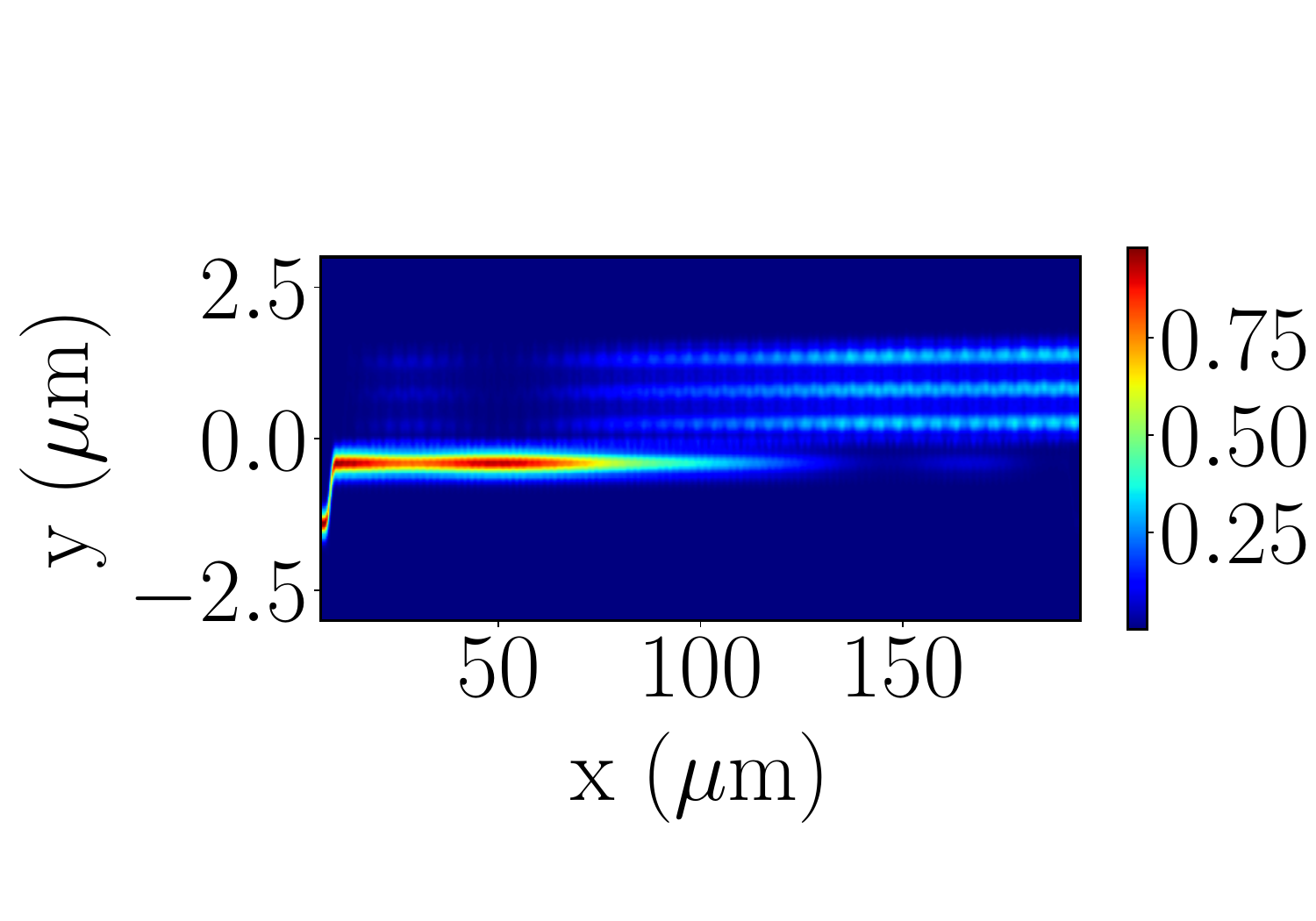}}\quad
\subfloat[\vspace{0pt}]{\includegraphics[width=0.3\columnwidth]{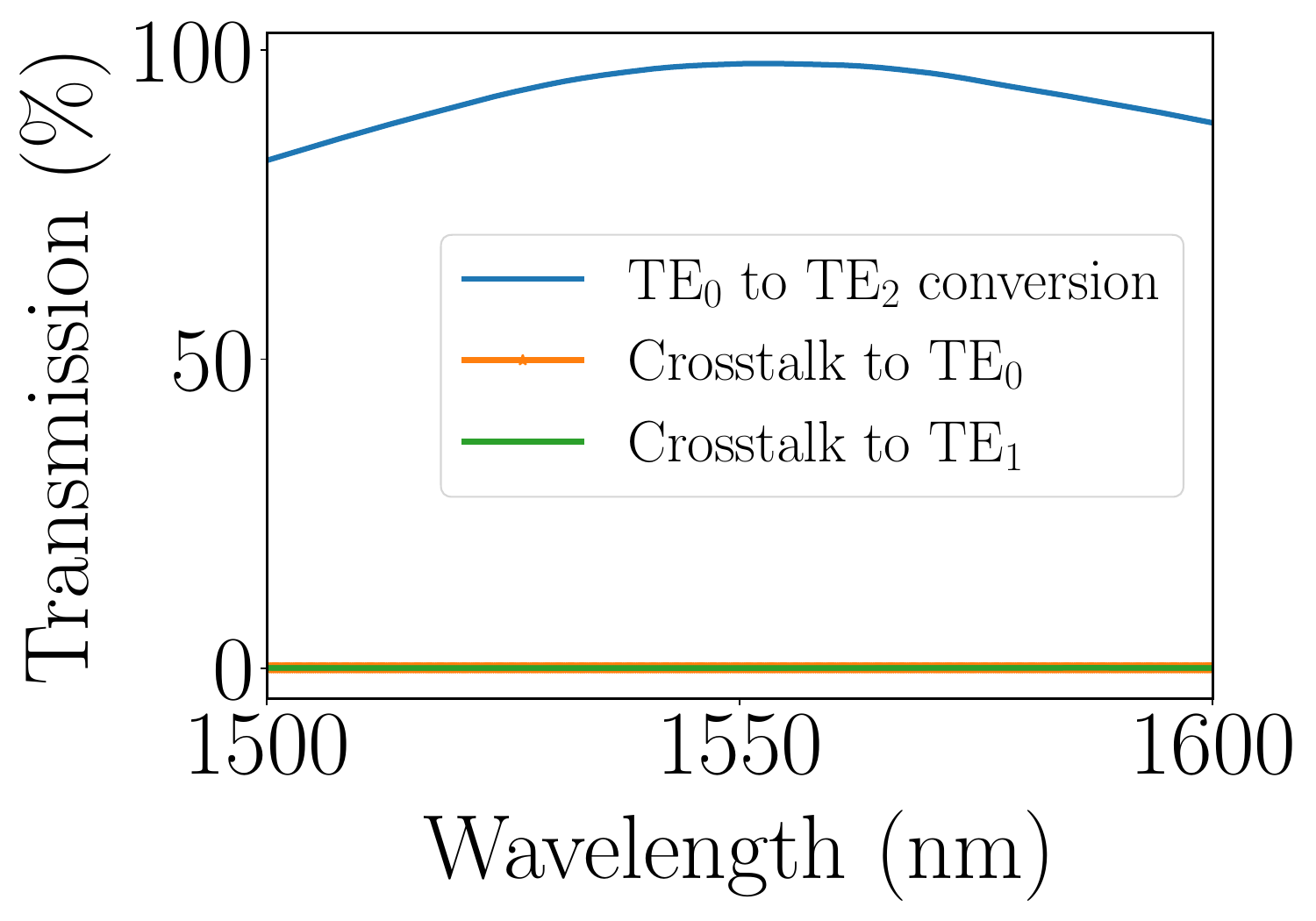}}\quad
\subfloat[\vspace{0pt}]{\includegraphics[width=0.3\columnwidth]{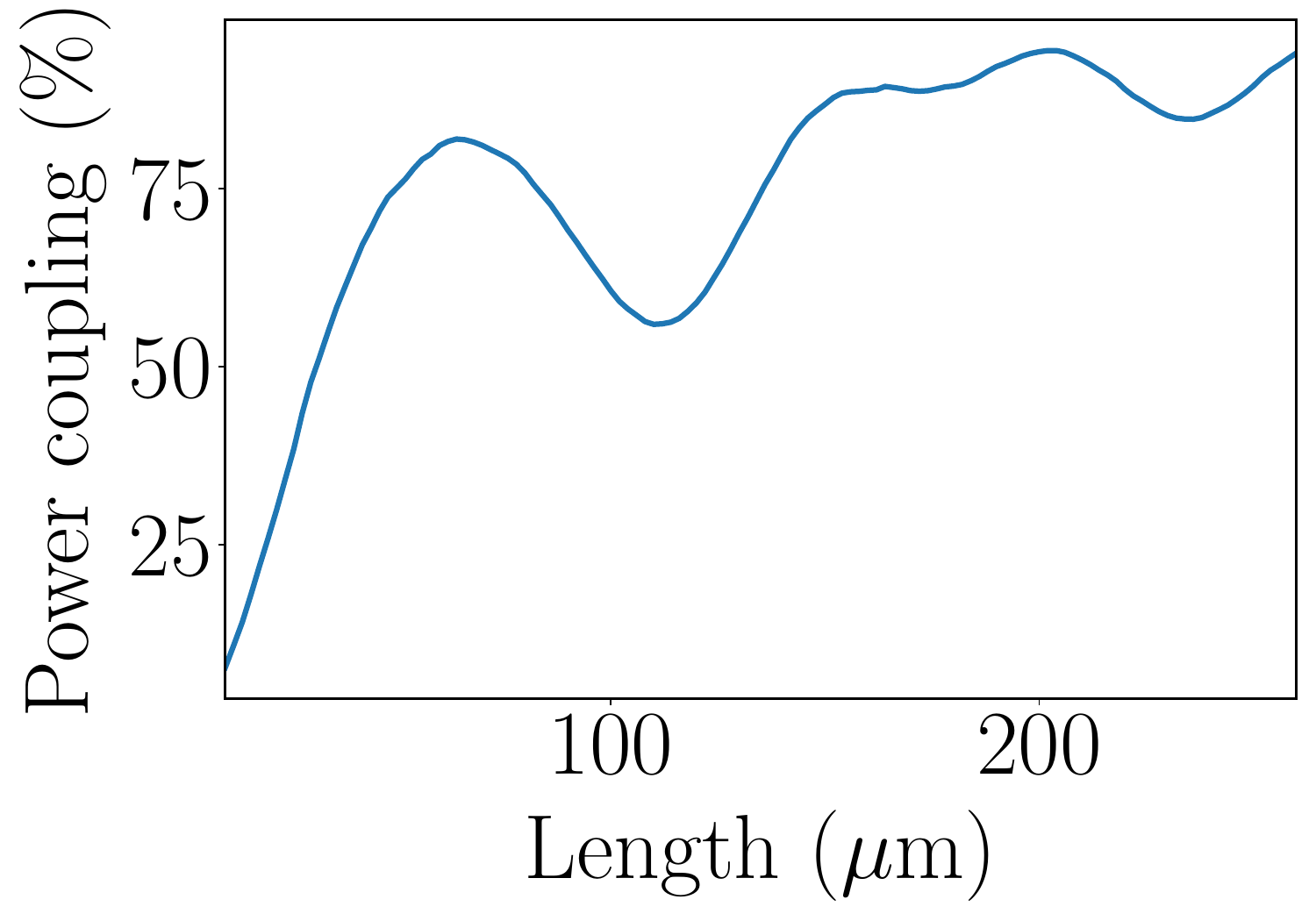}}\quad
\subfloat[\vspace{0pt}]{\includegraphics[width=0.3\columnwidth]{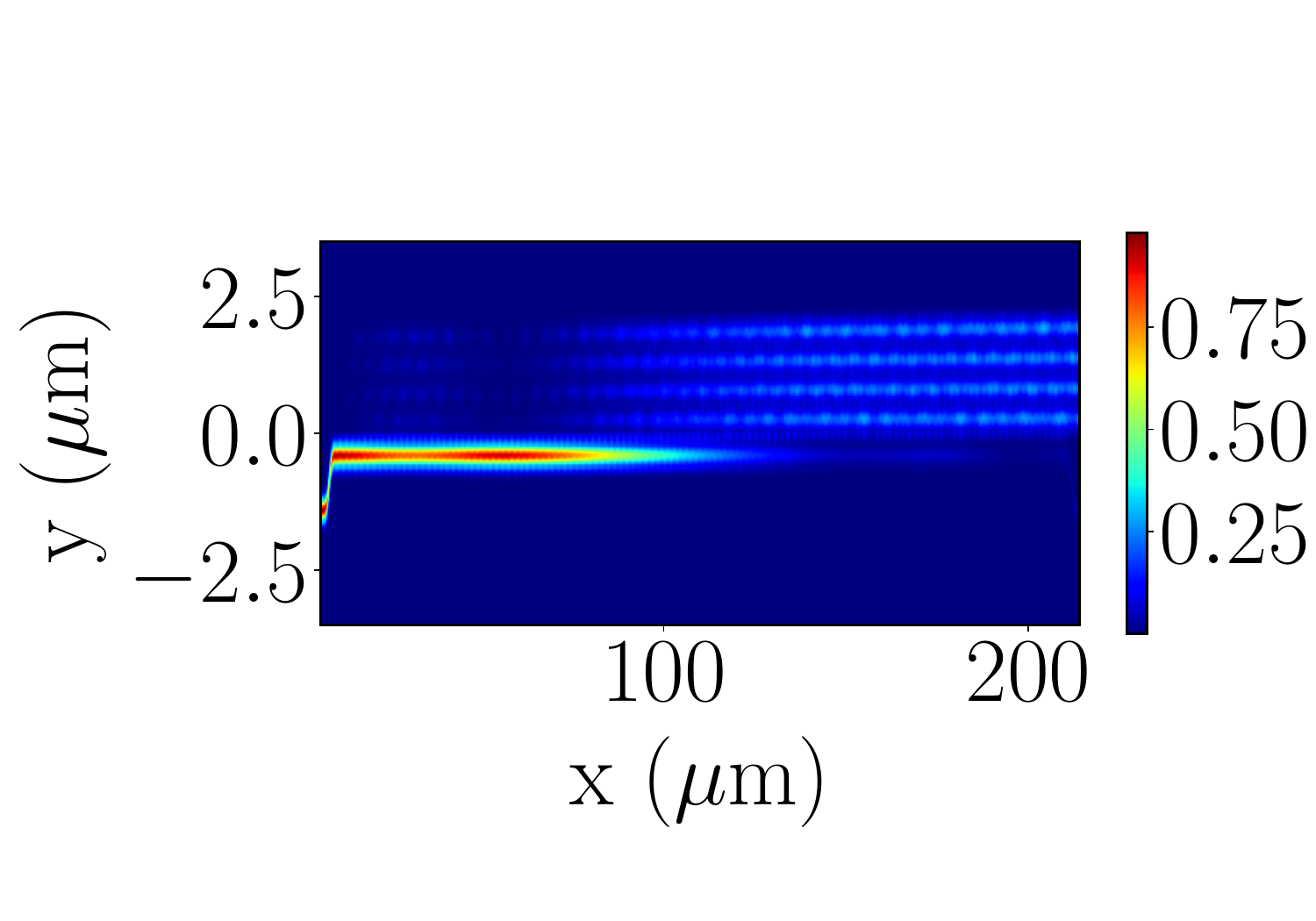}}\quad
\subfloat[\vspace{0pt}]{\includegraphics[width=0.3\columnwidth]{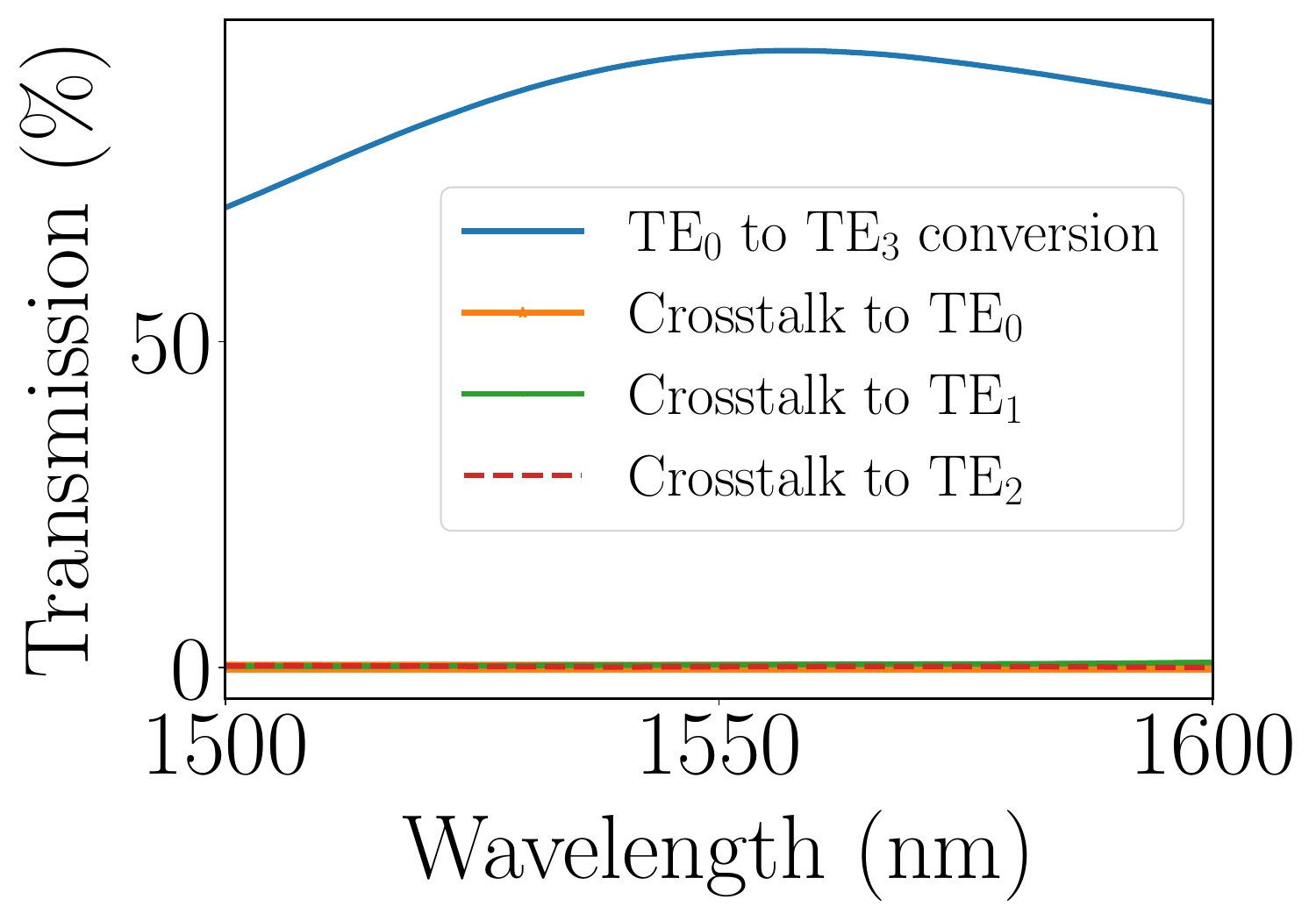}}\quad
\caption{(a), (d) and (g) Length optimization of $\mathrm{TE_1}$, $\mathrm{TE_2}$, and $\mathrm{TE_3}$ adiabatic DCs, respectively. (b), (e) and (h) Electric field intensity profile across $\mathrm{TE_1}$, $\mathrm{TE_2}$, and $\mathrm{TE_3}$ adiabatic DCs, respectively. (c), (f) and (i) Conversion efficiency and crosstalk in $\mathrm{TE_1}$, $\mathrm{TE_2}$, and $\mathrm{TE_3}$ adiabatic DCs, respectively.}
\label{fig:adc_sims}
\end{figure}

The fundamental-mode signals are multiplexed into higher-order modes in multimode waveguides using multimode adiabatic directional couplers (DC). The adiabatic DC is illustrated in Figure \ref{fig:adc}a. As per the coupled mode theory (CMT) \citesupp{pollock_coupled_2003-supp}, the coupling condition is achieved by optimizing the width of the multimode waveguide to match the effective index of the $\mathrm{TE_0}$ mode in the input waveguide with the desired mode effective index in the multimode waveguide. The adiabatic couplers are designed to have a variation in the multimode waveguide width, where the index matching point lies in the middle, as shown in Figure \ref{fig:adc}. The width of the multimode waveguide for index matching of each mode is chosen based on the index curves shown in Figure \ref{fig:adc}b, and the dimensions are given in Table \ref{Tb:adc}.

{
\begin{center}
\resizebox{0.475\textwidth}{!}{ 
\begin{tabular}{|c|c|c|c|c|}
\hline 
Mode &  $W_s$ ($\mu m$) &  $W_e$ ($\mu m$) & L ($\mu m$) & g ($nm$) \tabularnewline
\hline 
\hline 
$\mathrm{TE_1}$ & 0.96 & 1.08 & 160 & 170\tabularnewline
\hline
$\mathrm{TE_2}$ & 1.475 & 1.635 & 180 & 170\tabularnewline
\hline
$\mathrm{TE_3}$ & 1.985 & 2.185 & 200 & 170\tabularnewline
\hline
\end{tabular}
}
\end{center}
\captionof{table}{Adiabatic DC dimensions for $\mathrm{TE_0}$ to $\mathrm{TE_1}$-$\mathrm{TE_3}$ conversion.\label{Tb:adc}}
}

The adiabatic directional couplers (DCs) are optimized for maximum power coupling using the Lumerical Eigenmode Expansion (EME) software package \citesupp{lumerical-supp}. The optimization curves are shown in Figures \ref{fig:adc_sims}a, \ref{fig:adc_sims}d, and \ref{fig:adc_sims}g for $\mathrm{TE_1}$, $\mathrm{TE_2}$, and $\mathrm{TE_3}$ adiabatic DCs, respectively. As expected from CMT, the coupled power increases with the length of the coupler until it reaches saturation point. In addition, due to the adiabatic nature of the DCs, there is no power coupling back into the input arm. Therefore, the lengths of the couplers are selected to maximize coupling efficiencies and ensure diminishing reflected power. The optimized lengths are given in Table \ref{Tb:adc}. Figures \ref{fig:adc_sims}b, \ref{fig:adc_sims}e, and \ref{fig:adc_sims}h demonstrate the successful mode conversion and power propagation from $\mathrm{TE_0}$ to $\mathrm{TE_1}$, $\mathrm{TE_2}$ and $\mathrm{TE_3}$ in the adiabatic DCs, respectively. Additionally, Figures \ref{fig:adc_sims}c, \ref{fig:adc_sims}f, and \ref{fig:adc_sims}i illustrate conversion efficiency and crosstalk to other modes across wavelength for $\mathrm{TE_1}$, $\mathrm{TE_2}$, and $\mathrm{TE_3}$ adiabatic DCs, respectively.

\section{Multimode microring resonators}\label{app:mm_ring}
\begin{figure}[hbt]
    \centering
    \includegraphics[width=0.5\columnwidth]{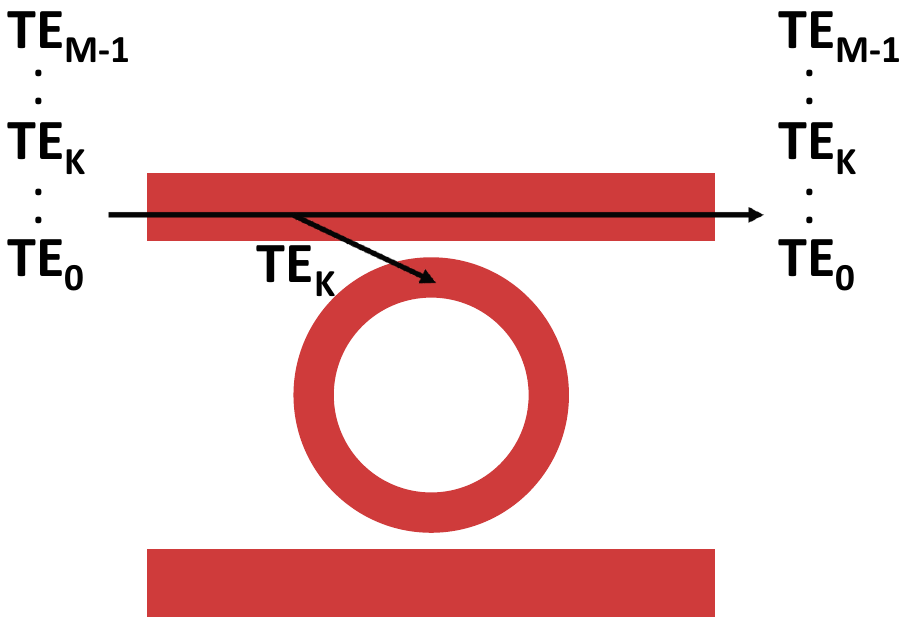}
    \caption{MDM-WDM microring resonator. Mode selectivity is controlled by the bus waveguide width and wavelength selectivity is controlled by the ring radius.}
    \label{fig:mdm_ring}
\end{figure}

\begin{figure}[h!bt]
\subfloat[\vspace{0pt}]{\includegraphics[width=0.49\columnwidth]{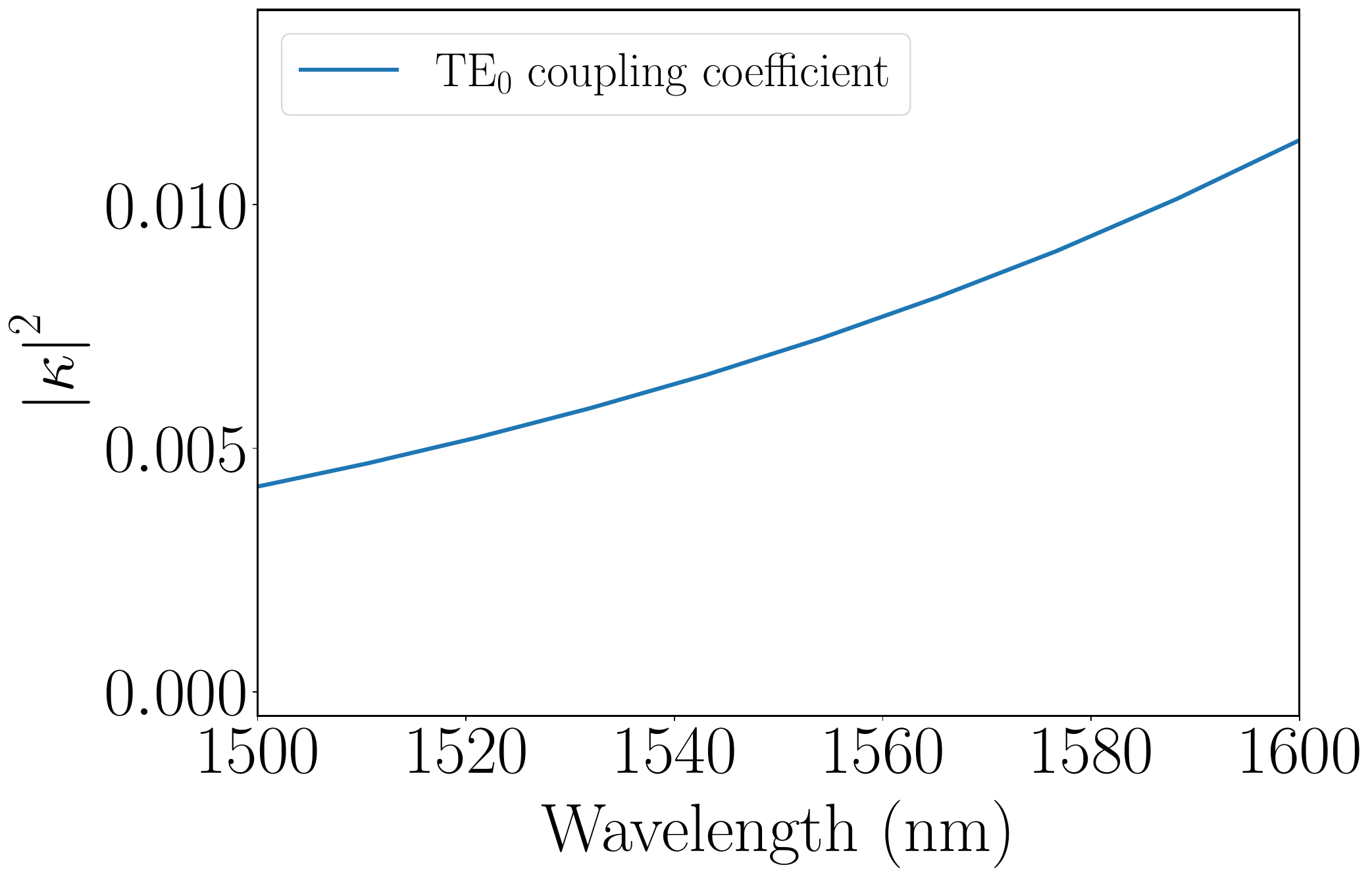}}\quad
\subfloat[\vspace{0pt}]{\includegraphics[width=0.49\columnwidth]{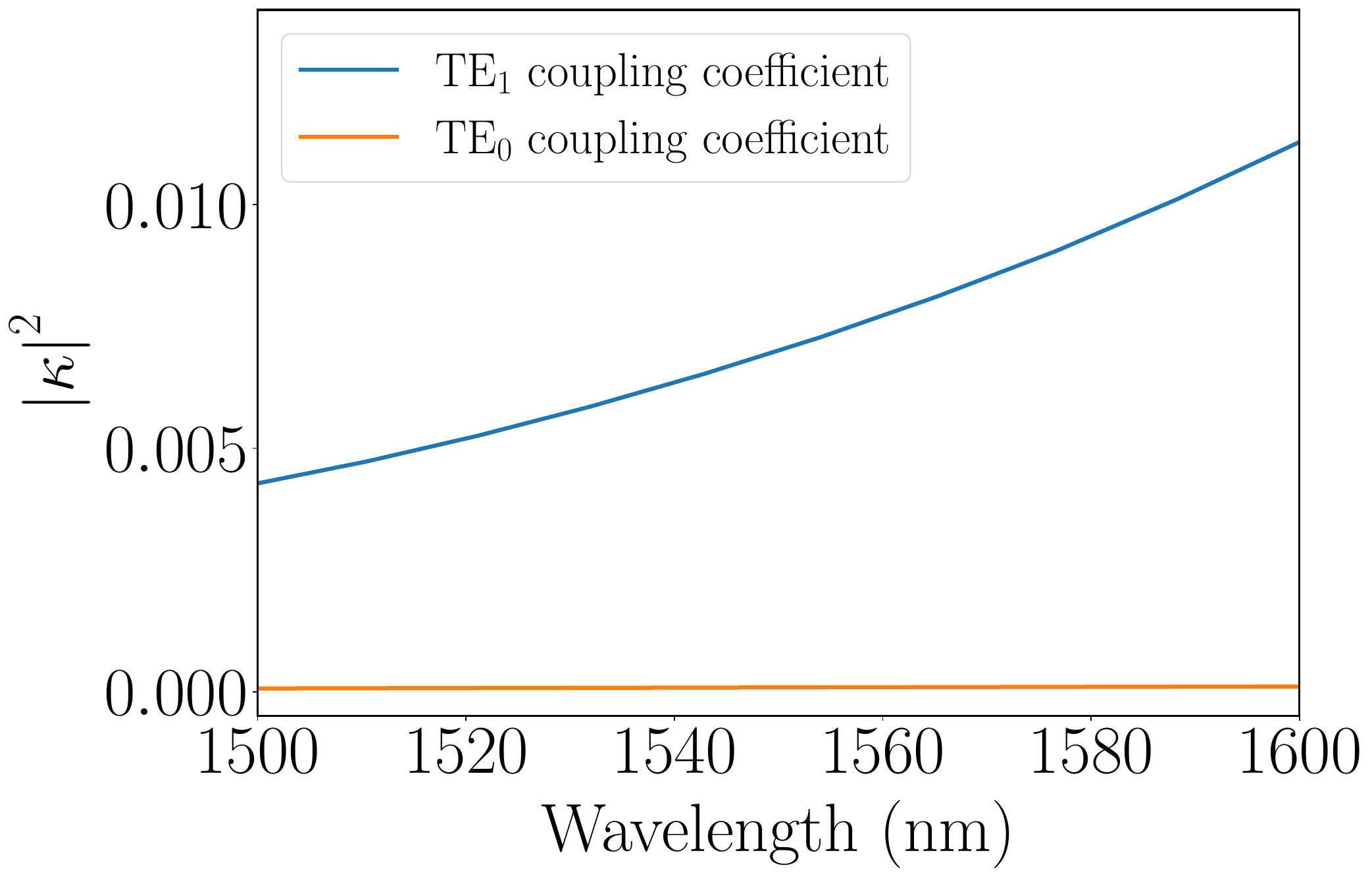}}\quad
\subfloat[\vspace{0pt}]{\includegraphics[width=0.49\columnwidth]{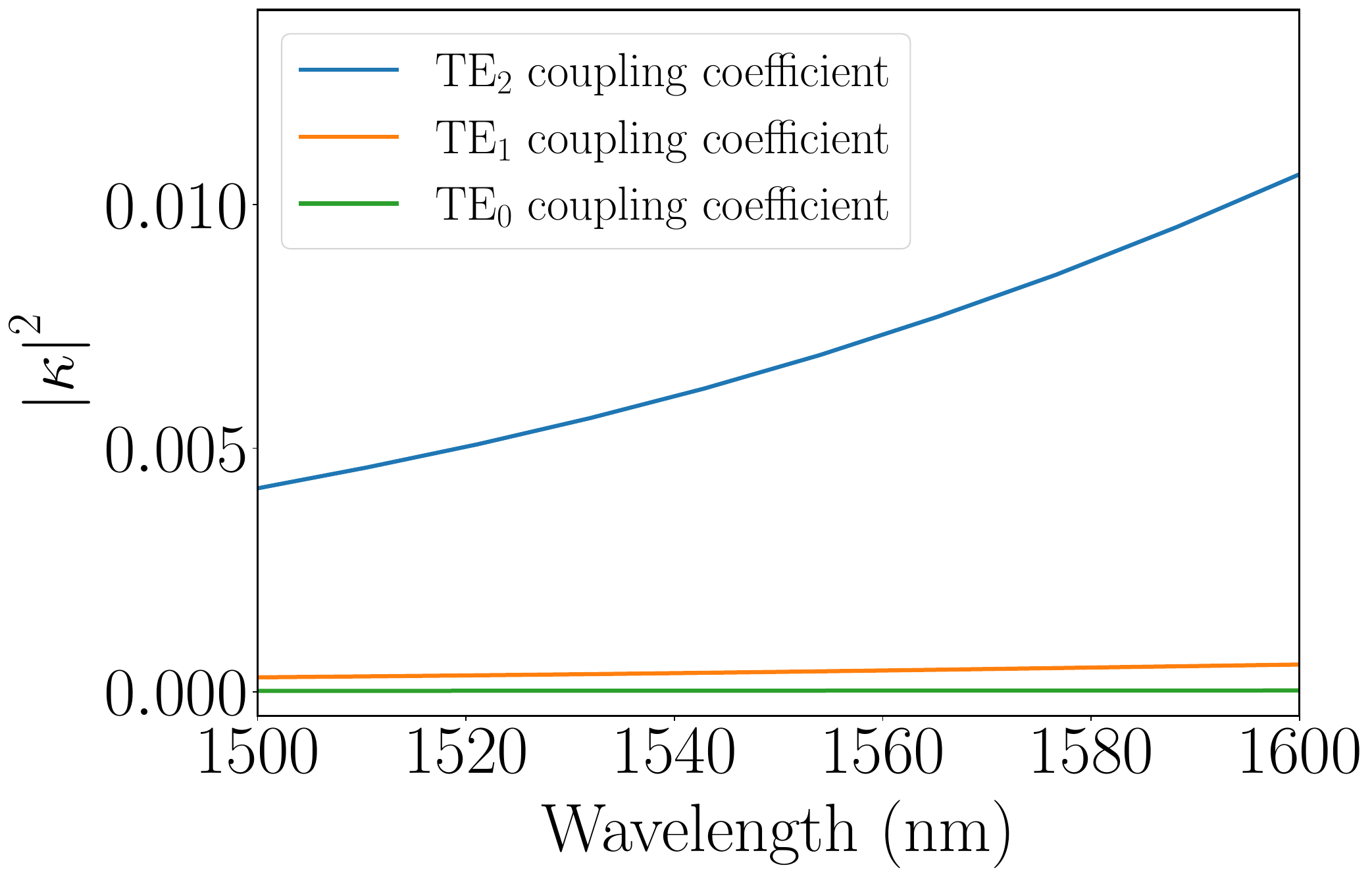}}\quad
\subfloat[\vspace{0pt}]{\includegraphics[width=0.49\columnwidth]{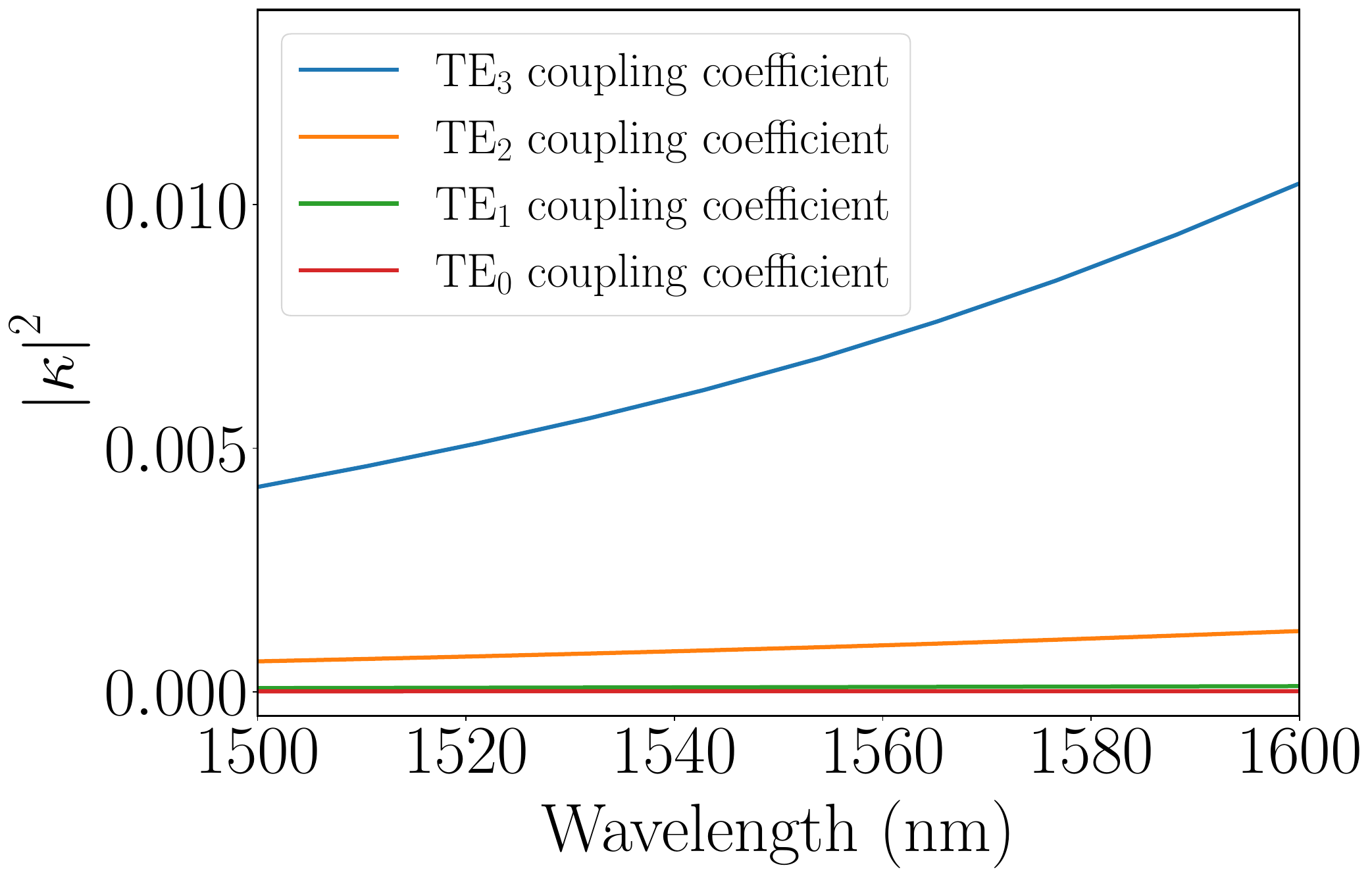}}
\caption{Simulated power coupling coefficient $\left| \kappa \right|^2$ of the ring point couplers for the allowed modes in the bus waveguide for (a) $\mathrm{TE_0}$ ring, (b) $\mathrm{TE_1}$ ring, (c) $\mathrm{TE_2}$ ring, and (d) $\mathrm{TE_3}$ ring. It is evident that the ring couplers only interact with the correct mode.}
\label{fig:mdm_ring_couplers}
\end{figure}

The mode-divison multiplexing (MDM) and wavelength-division multiplexing (WDM) processor requires the ability to be mode-selective as well as being wavelength-selective. The mode and wavelength-selective microring resonator (MRR) is shown in Figure \ref{fig:mdm_ring}. The multimode MRR resonator is composed of a single-mode microring waveguide and a multimode bus waveguide. This MRR is mode-selective, because the index of the higher-order mode in the bus waveguide has to match the $\mathrm{TE_0}$ index in the single-mode waveguide of the ring by tuning the width of the bus waveguide. This MRR is also wavelength-selective due to resonance wavelength condition, achieved by tuning its radius as follows:

\begin{equation}
    \lambda_{res} = \frac{2 \pi R n_{eff}}{m},
    \label{eq:resonance}
\end{equation}
where $\lambda_{res}$ is the resonance wavelength, $R$ is the ring radius and $n_{eff}$ is the mode effective index inside the ring waveguide, and $m$ is the resonant mode order.

{
\begin{center}
\resizebox{0.25\textwidth}{!}{ 
\begin{tabular}{|c|c|c|}
\hline 
Mode  & $W_{bus}$ ($\mu m$) 
\tabularnewline
\hline 
\hline 
$\mathrm{TE_0}$ & 0.5 \tabularnewline
\hline
$\mathrm{TE_1}$ & 1.02 \tabularnewline
\hline
$\mathrm{TE_2}$ & 1.556\tabularnewline
\hline
$\mathrm{TE_3}$ & 2.084\tabularnewline
\hline
\hline
$\lambda_{res}$ (nm) & $R$ ($\mu m$) \tabularnewline
\hline
1554.387 & 8 \tabularnewline
\hline
1557.588 & 8.02843 \tabularnewline
\hline
\end{tabular}
}
\end{center}
\captionof{table}{The bus widths of the microring resonators required for interaction with each mode, and the radii required for interaction with each wavelength.\label{Tb:mdm_rings}}
}

$\mathrm{TE_0}$ to $\mathrm{TE_3}$ mode and wavelength-selective rings are designed based on the mode index curves in Figure \ref{fig:adc}b and the resonance wavelength condition in equation \ref{eq:resonance}. The design values are listed in Table \ref{Tb:mdm_rings}. The ring point couplers were simulated to extract the power coupling coefficient for each mode, as illustrated in Figure \ref{fig:mdm_ring_couplers}. As shown in Figure \ref{fig:mdm_ring_couplers}, each ring coupler interacts strongly with the correct mode, thus ensuring that the rings are mode-selective.

\section{Multimode photodetectors}
\label{sec:detectors}

\begin{figure}[hbt]
    \centering
    \includegraphics[width=0.45\columnwidth]{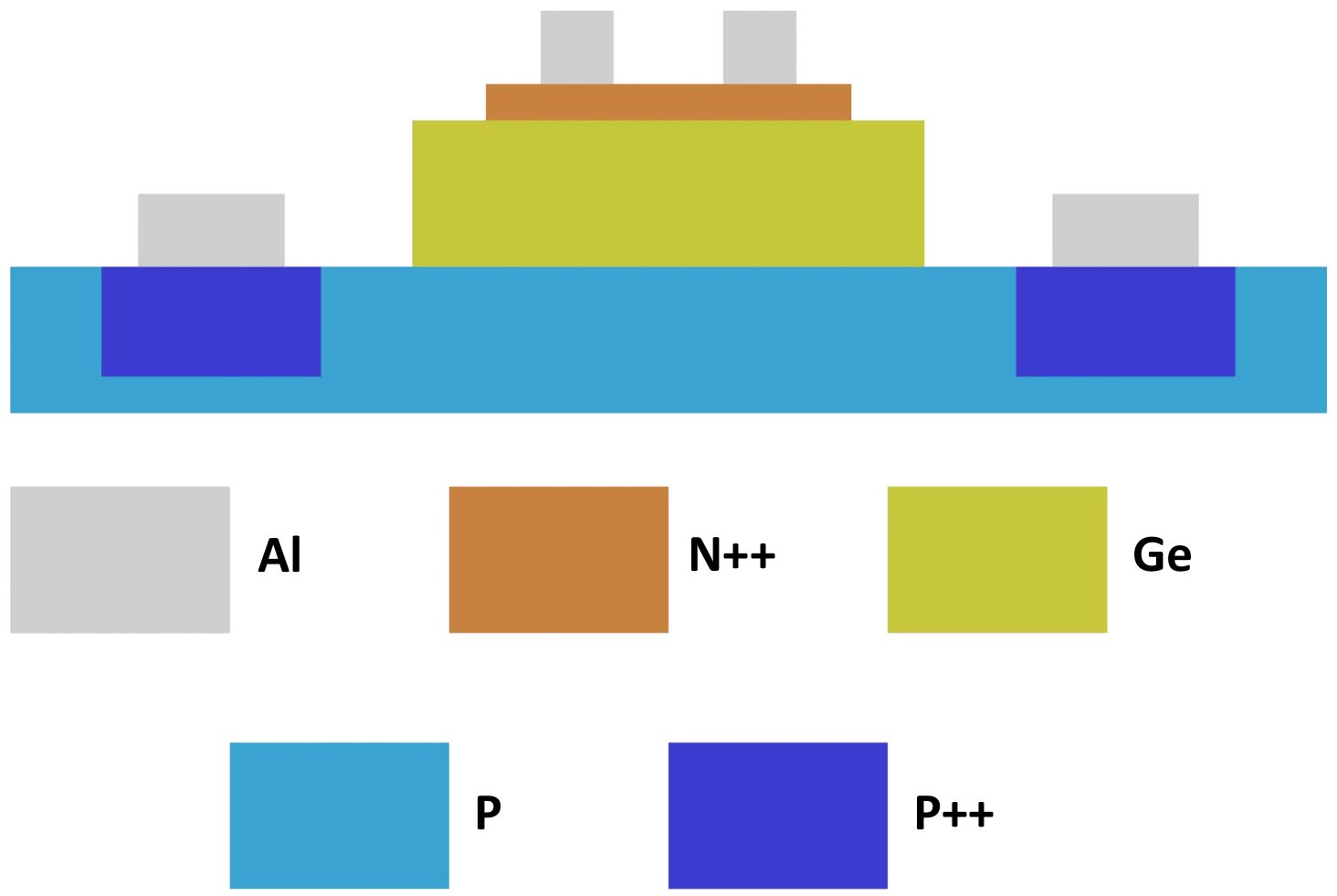}
    \caption{Cross-section of the multimode germanium PIN vertical photodetector.}
    \label{fig:xsec_PD}
\end{figure}

\begin{figure}[h!bt]
\subfloat[\vspace{0pt}]{\label{fig:PD_TE0_power}\includegraphics[width=0.45\columnwidth]{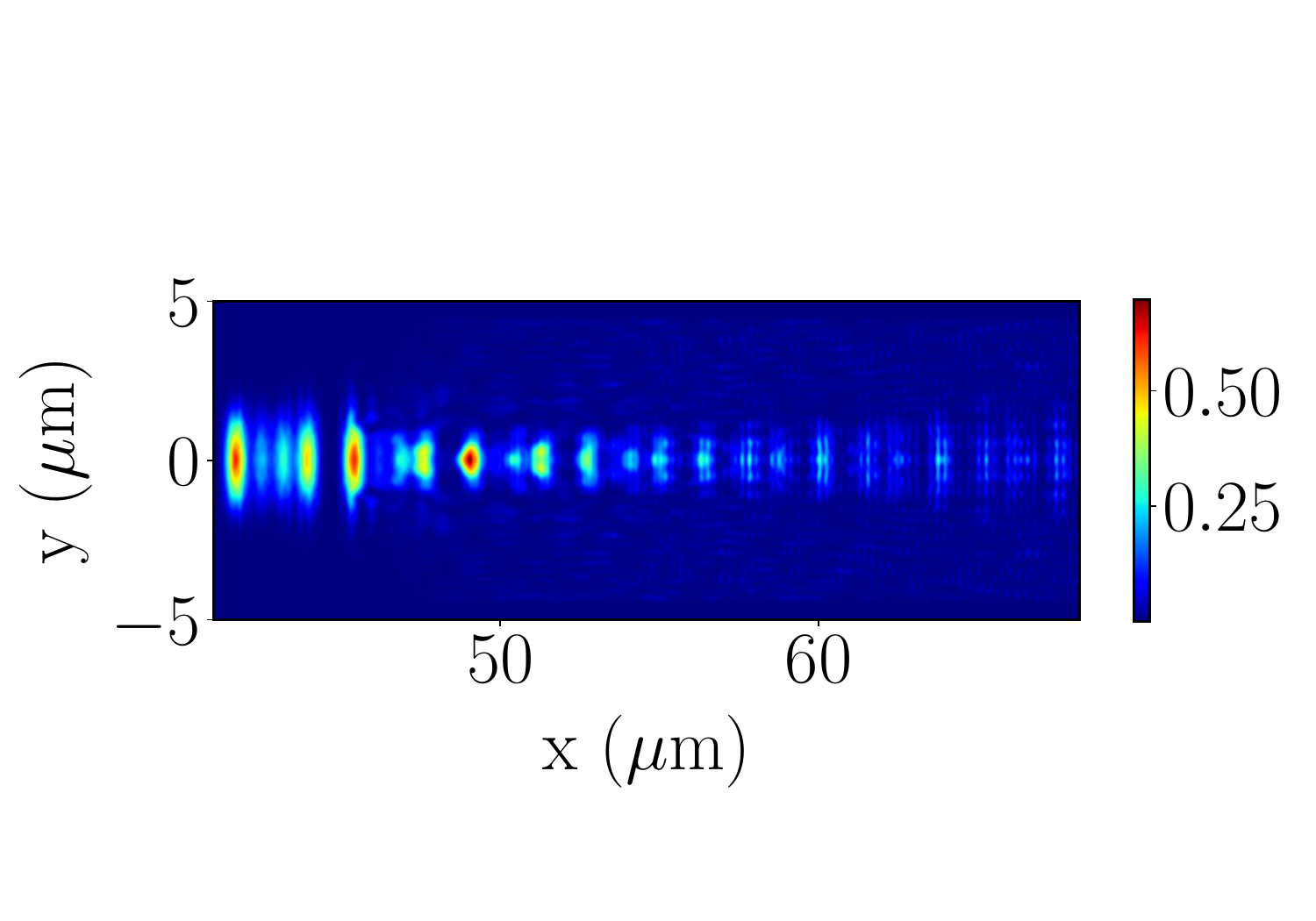}}\quad
\subfloat[\vspace{0pt}]{\label{fig:PD_TE1_power}\includegraphics[width=0.45\columnwidth]{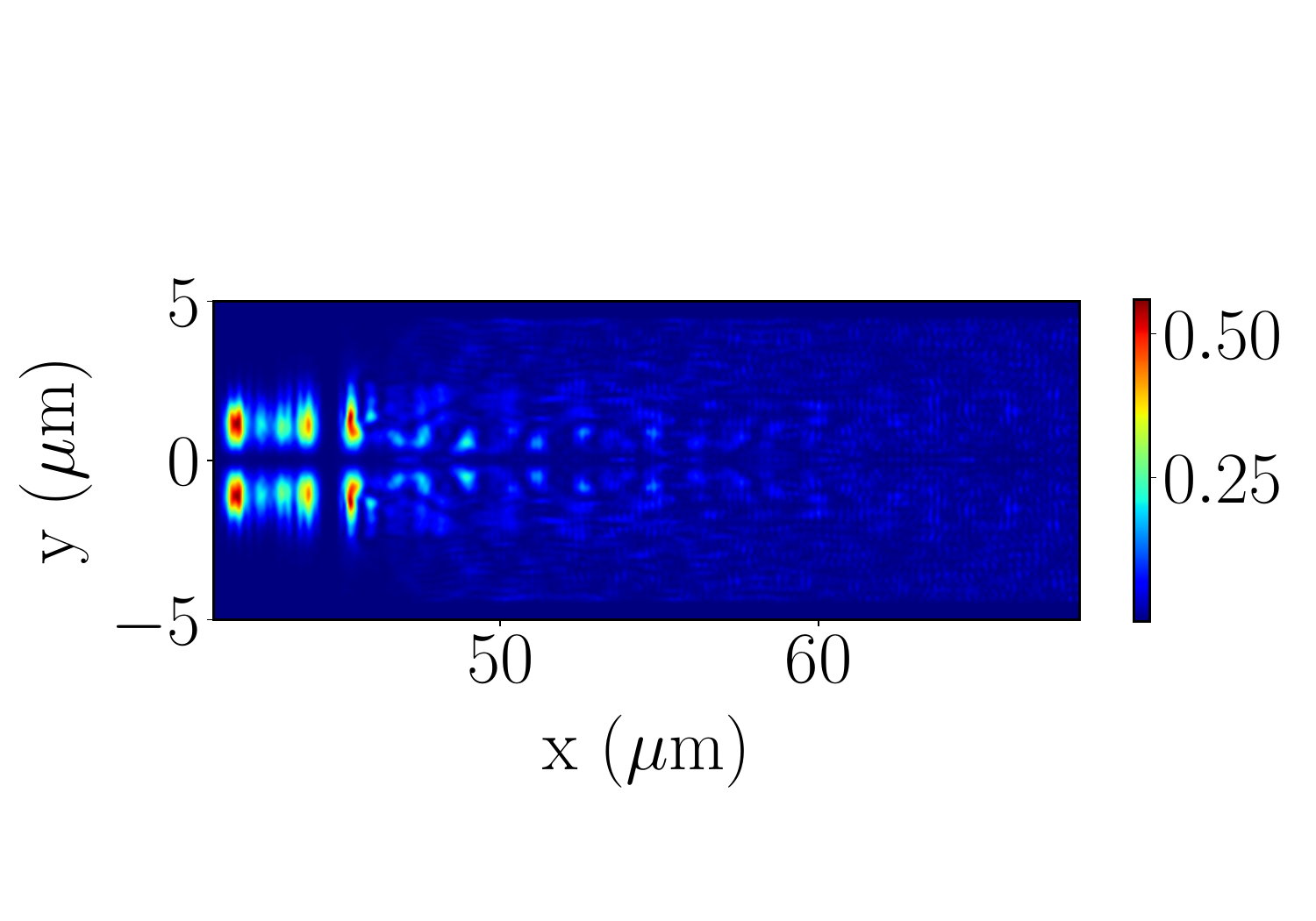}}\quad
\subfloat[\vspace{0pt}]{\label{fig:PD_TE2_power}\includegraphics[width=0.45\columnwidth]{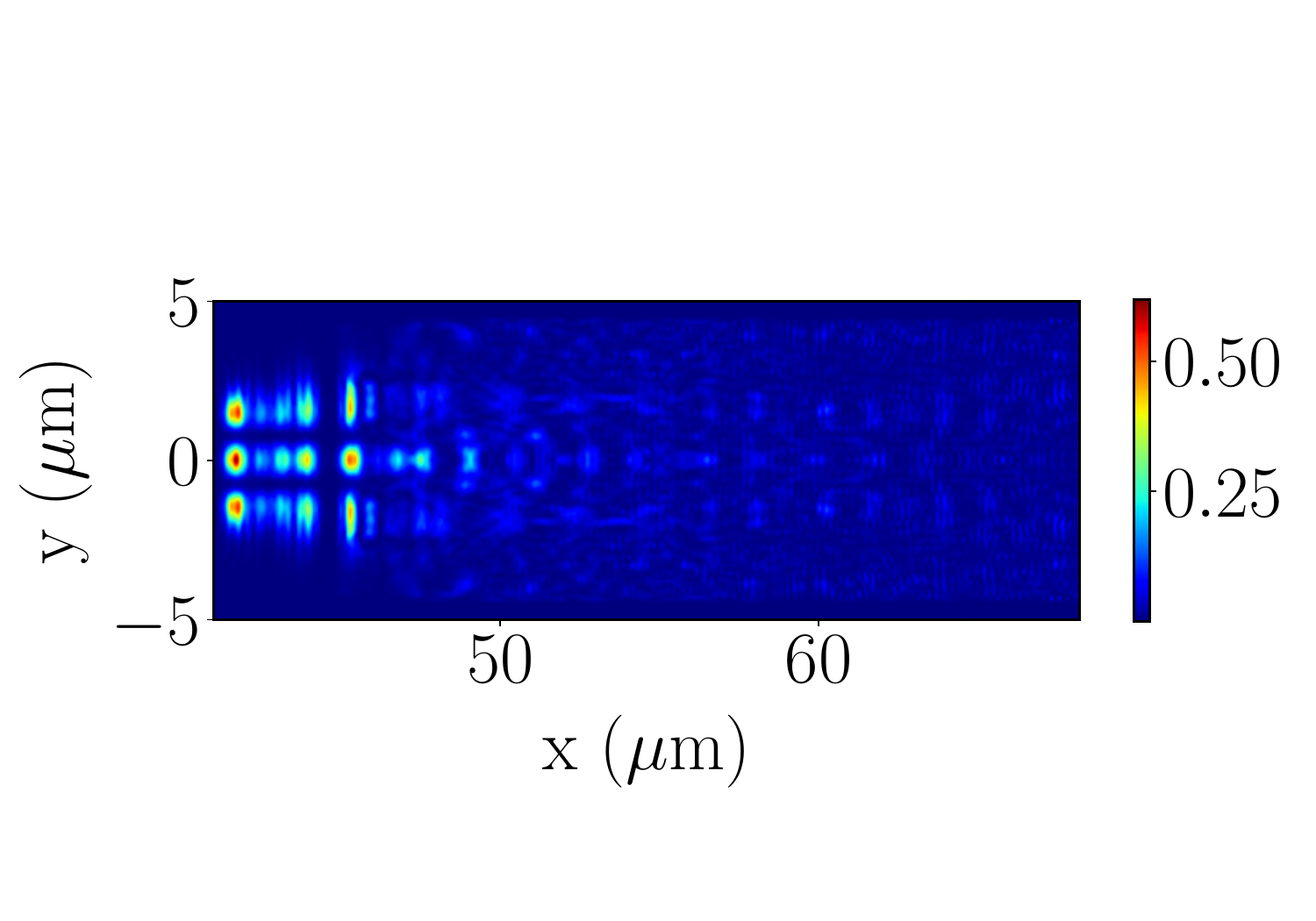}}\quad
\subfloat[\vspace{0pt}]{\label{fig:PD_TE3_power}\includegraphics[width=0.45\columnwidth]{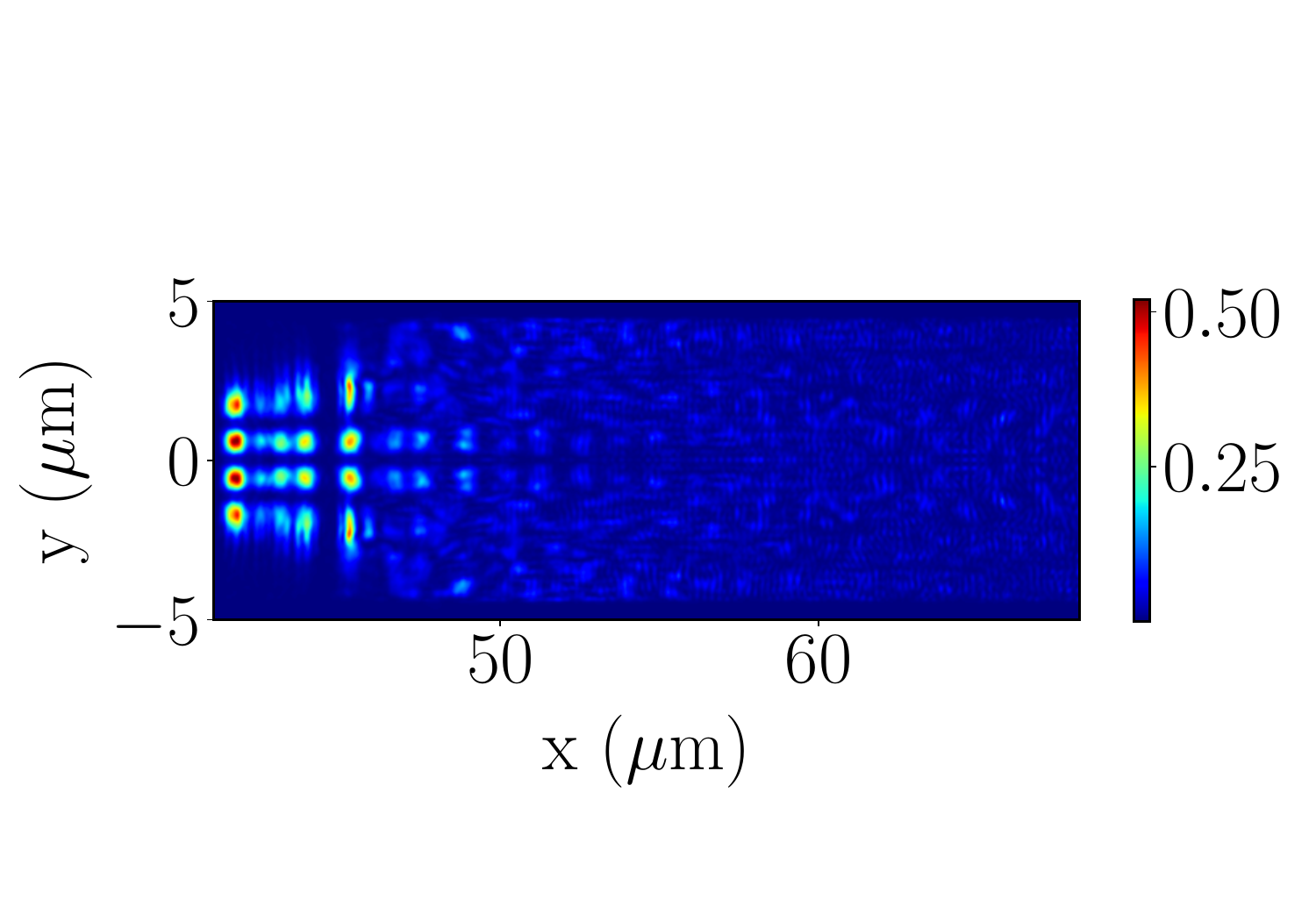}}
\caption{Electric field intensity profile in the germanium layer of the photodetector for (a) $\mathrm{TE_0}$ mode, (b) $\mathrm{TE_1}$ mode, (c) $\mathrm{TE_2}$ mode, and (d) $\mathrm{TE_3}$ mode.}
\label{fig:ge_power}
\end{figure}

The addition of weighted multimode optical signals is performed at the multimode photodetectors via carrier accumulation. We integrated a PIN germanium vertical photodetector (PD) at the end of our processor. Our final multimode waveguide width at the photodetector is chosen to be 2.185 $\mu$m to support $\mathrm{TE_0}$ to $\mathrm{TE_3}$ modes and to index-match the final adiabatic DC width. Additionally, we split the top metal contacts connecting to the N-doped section of the PD into two sections. Therefore the metal contacts are far from the location of the peak power for $\mathrm{TE_0}$ to $\mathrm{TE_3}$ modes, resulting in enhancement in the PD responsivity \citesupp{fard_responsivity_2016-supp}. The cross-section of the PD is shown in Figure \ref{fig:xsec_PD}. The intensity profiles of the electric field coupled to the germanium layer of the PD for $\mathrm{TE_0}$ to $\mathrm{TE_3}$ modes are extracted using Lumerical finite-difference time-domain method (FDTD) software packages, and are shown in Figure \ref{fig:ge_power}. The responsivity and eye diagrams of $\mathrm{TE_0}$ to $\mathrm{TE_3}$ modes for the fabricated PD are reported in Figure \ref{fig:device} in the manuscript.

\section{Experimental setup}\label{app:setup}

\begin{figure}[hbt]
    \centering
    \includegraphics[width=1\columnwidth]{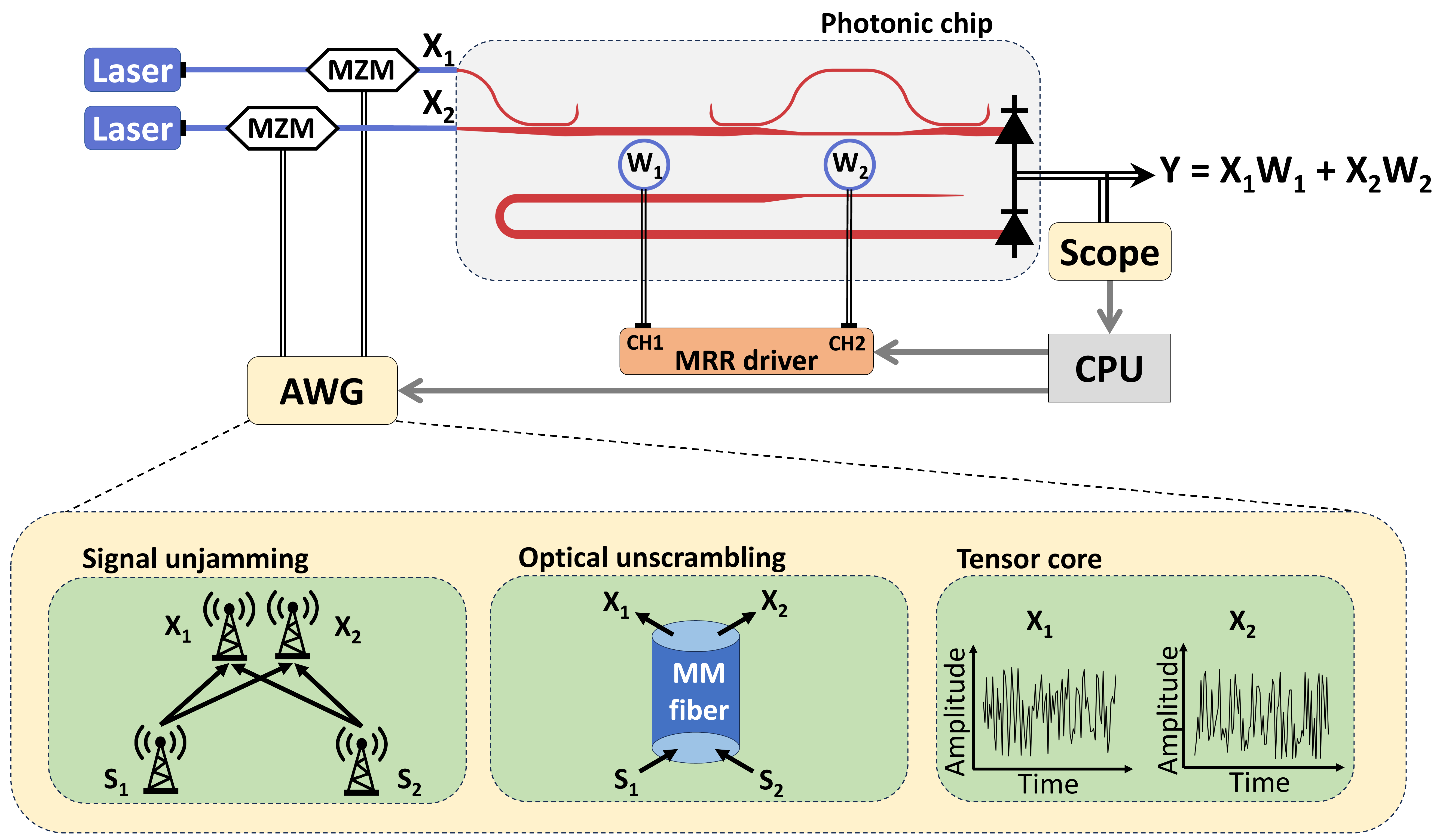}
    \caption{Diagram illustrating the signal processing experimental setup.}
    \label{fig:mdm_rf_setup}
\end{figure}

We demonstrate two different applications by configuring two arbitrary waveform generator (AWG) channels to emulate the scrambling of two optical bit streams in a multimode fiber or mixing a PSK signal with a jammer signal in RF multiple-input and multiple-output (MIMO) systems. Two lasers of the same wavelength (1530.6 nm) are modulated by Mach-Zehnder modulators (MZM) that are driven by the AWG signals as shown in Figure \ref{fig:mdm_rf_setup}. The input light signals are sent into the $\mathrm{TE_0}$ and $\mathrm{TE_1}$ channels of the multimode processor. The multimode MRR photonic weights are configured depending on the demonstrated signal processing application, such as unscrambling optical streams or signal unjamming. The PDs are reverse-biased by 3V using a bias tee, and the output signal is monitored using a high-speed scope. 

\begin{figure}[bht]
\subfloat[\vspace{0pt}]{\label{fig:lut_te0}\includegraphics[width=0.49\columnwidth]{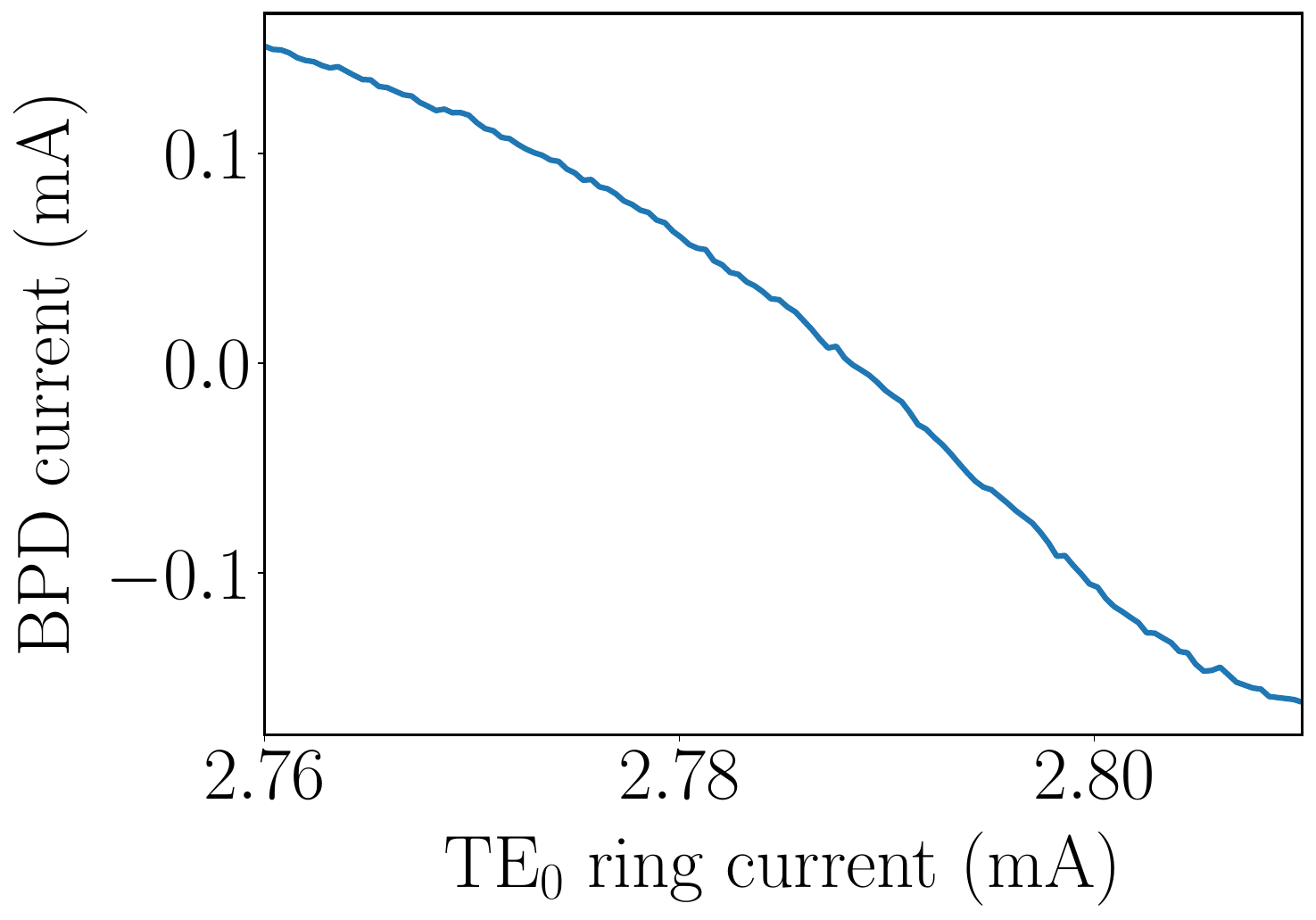}}\quad
\subfloat[\vspace{0pt}]{\label{fig:lut_te1}\includegraphics[width=0.49\columnwidth]{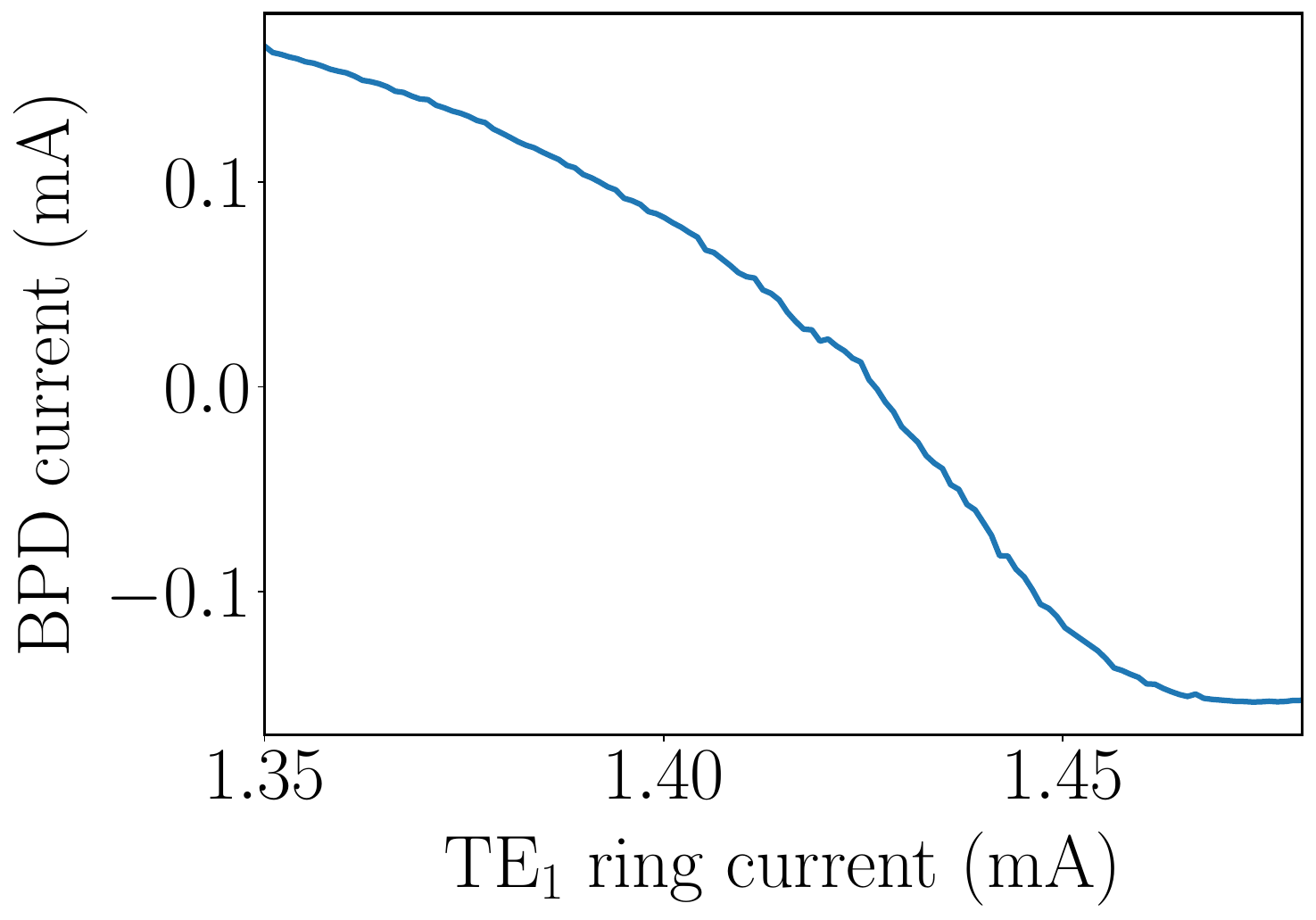}}
\caption{(a) The extracted LUT for the $\mathrm{TE_0}$ weight. (b) The extracted LUT for the $\mathrm{TE_1}$ weight.}
\label{fig:luts}
\end{figure}

\begin{figure}[bht]
\subfloat[\vspace{0pt}]{\label{fig:thermal_xtalk_TE0}\includegraphics[width=0.49\columnwidth]{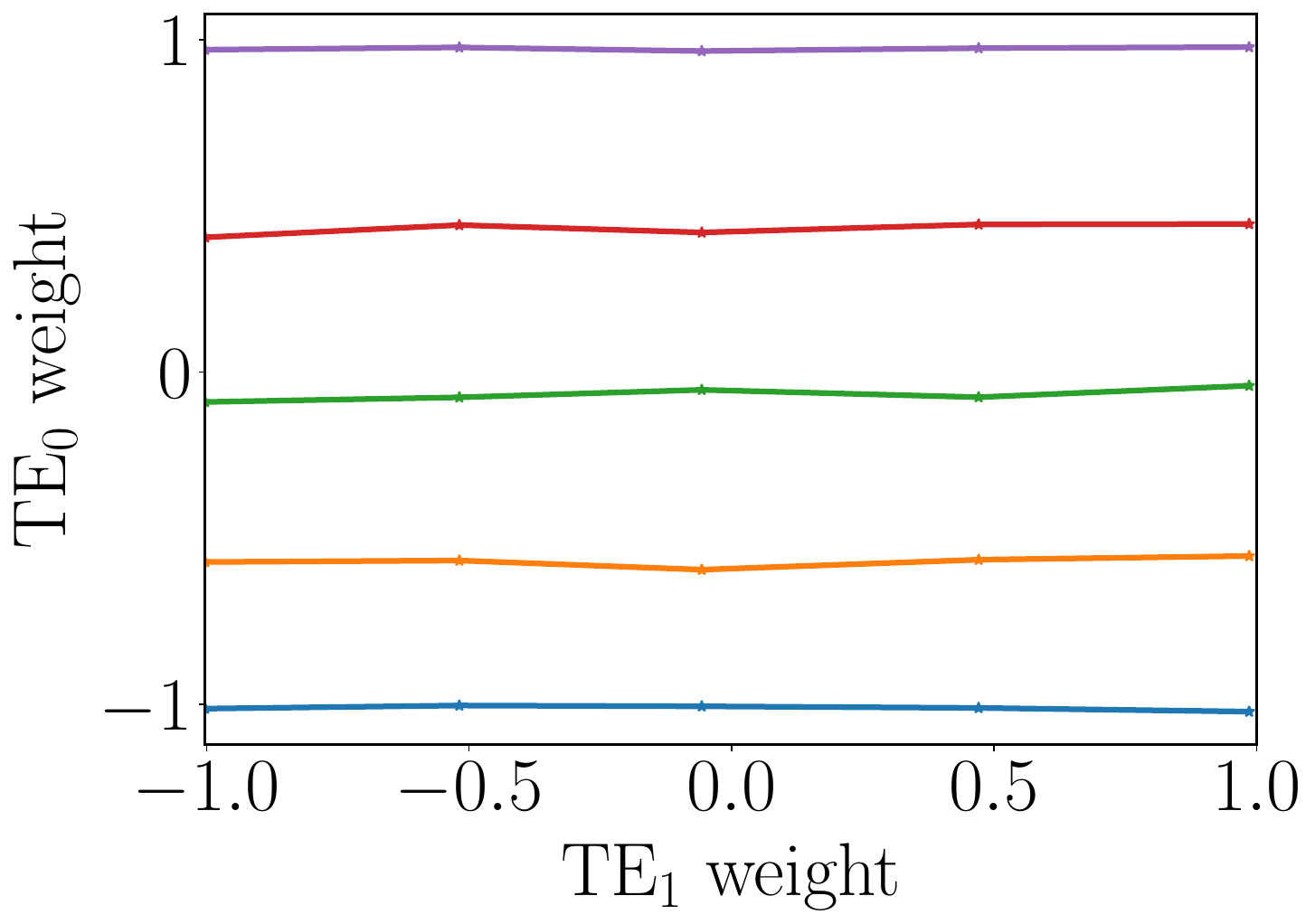}}\quad
\subfloat[\vspace{0pt}]{\label{fig:thermal_xtalk_TE1}\includegraphics[width=0.49\columnwidth]{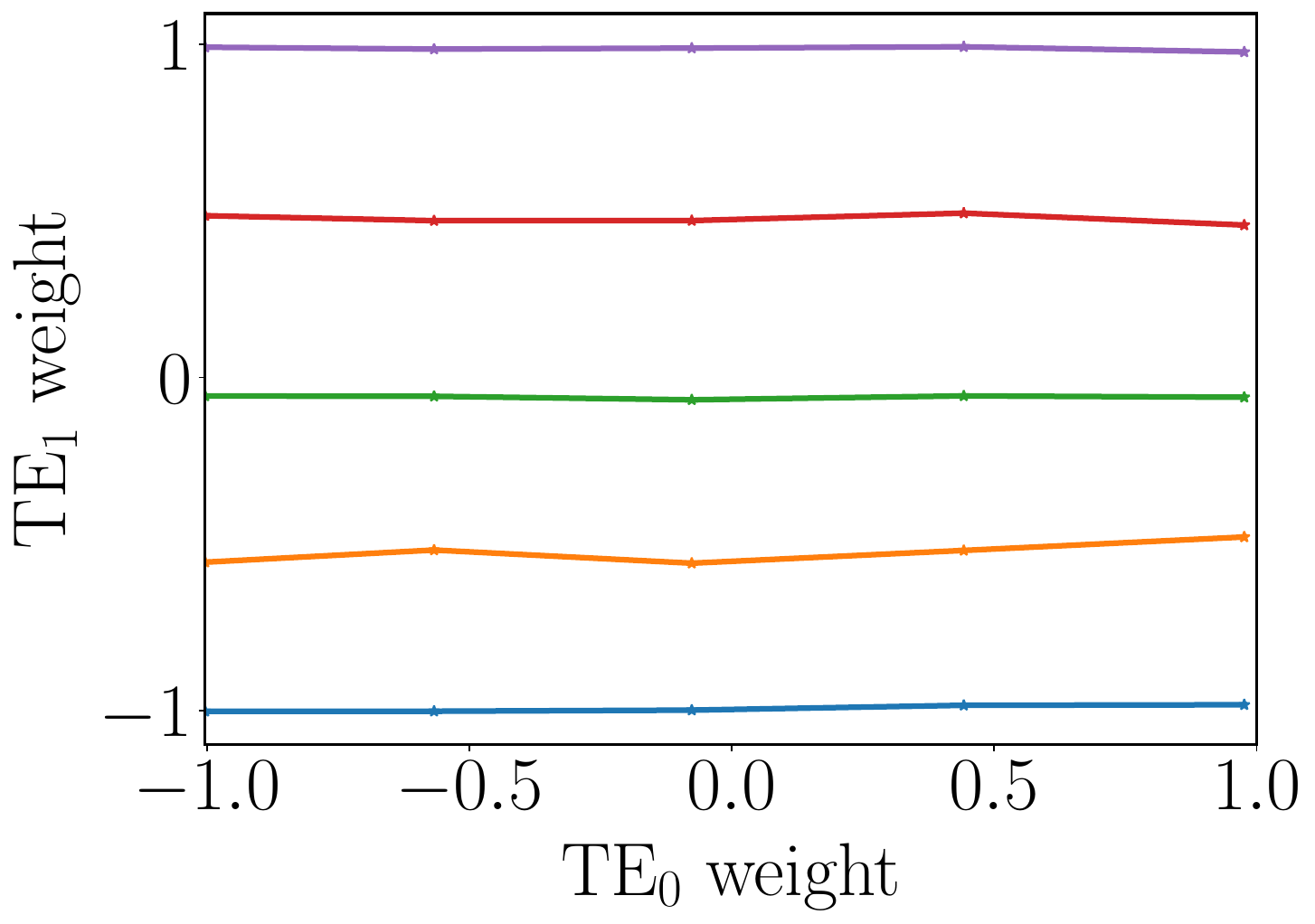}}
\caption{(a) Crosstalk to $\mathrm{TE_0}$ weight due to $\mathrm{TE_1}$ weight. Multiple $\mathrm{TE_0}$ weights are monitored while sweeping $\mathrm{TE_1}$ weight across its entire range. (b) Crosstalk to $\mathrm{TE_1}$ weight due to $\mathrm{TE_0}$ weight. Multiple $\mathrm{TE_1}$ weights are monitored while sweeping $\mathrm{TE_0}$ weight across its entire range. We observe very little inter-weight crosstalk.}
\label{fig:thermal_xtalk}
\end{figure}

In order to configure the MRRs, we calibrate our weights to obtain the heater current to the weight look-up tables (LUT). The extracted LUTs for $\mathrm{TE_0}$ and $\mathrm{TE_1}$ modes are shown in Figures \ref{fig:lut_te0} and \ref{fig:lut_te1}, respectively. It is noteworthy that we barely notice any thermal crosstalk between weights, as shown in Figure \ref{fig:thermal_xtalk}. As a result, we can use the 1D LUTs instead of 2D calibration, drastically reducing the calibration time.

\section{WDM-SMS and WDM-MDM photonic processor}\label{app:sms}

\begin{figure}[bht]
\centering
\subfloat[\vspace{0pt}]{\label{fig:scale-sms}\includegraphics[width=0.8\columnwidth]{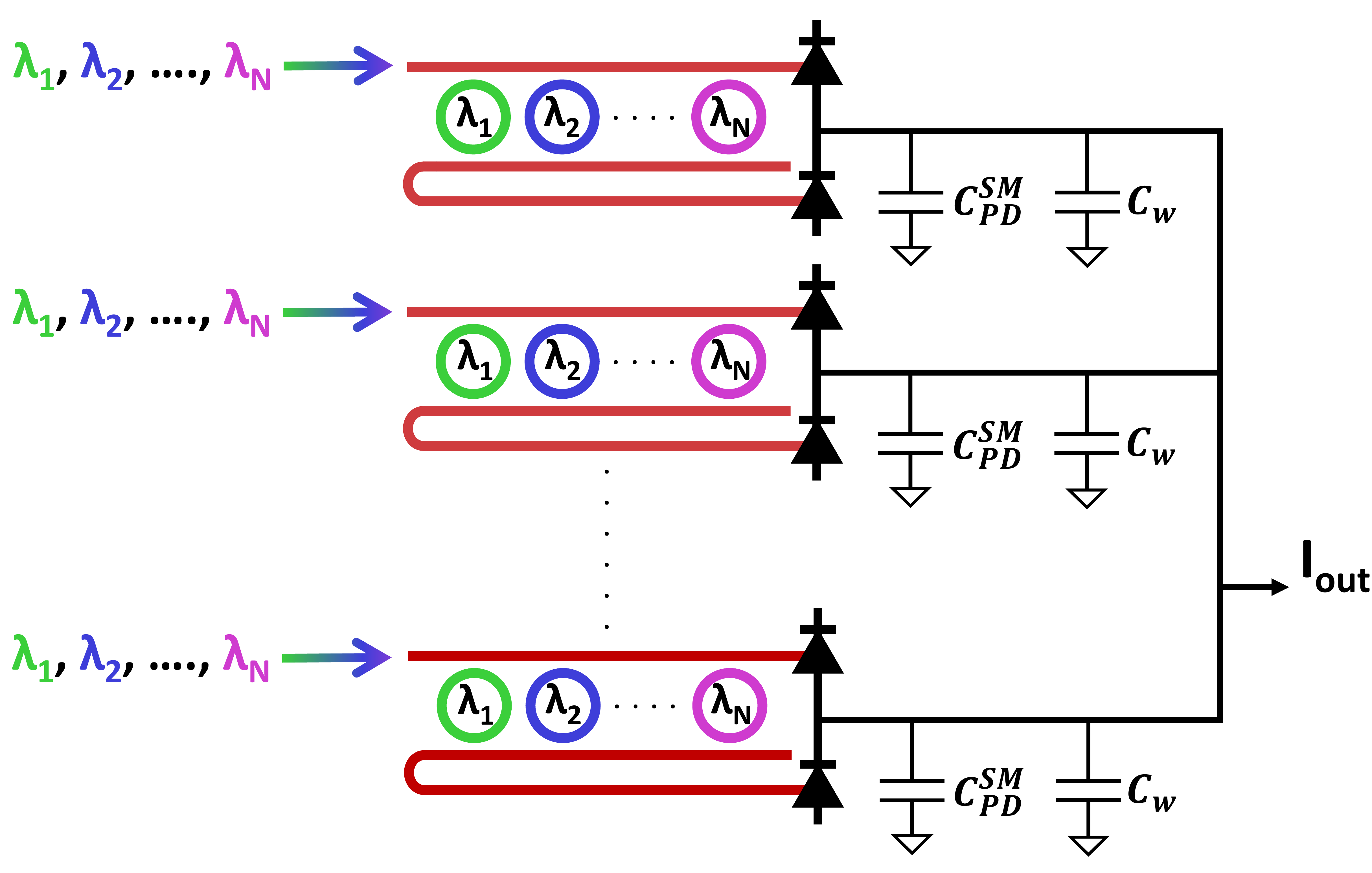}}\quad
\subfloat[\vspace{0pt}]{\label{fig:scale-mdm}\includegraphics[width=1\columnwidth]{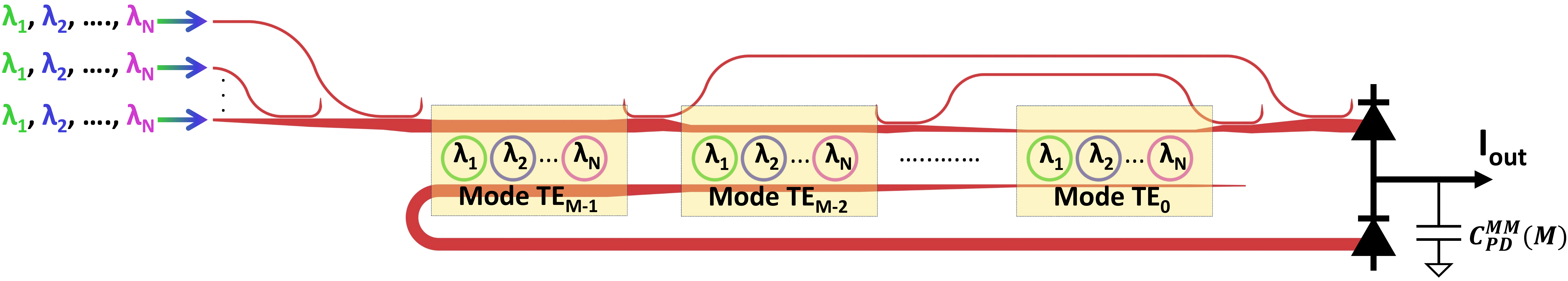}}
\caption{Comparison between the (a) WDM-SMS and (b) WDM-MDM photonic processor. M is the number of modes.}
\label{fig:SMS}
\end{figure}

When multiple WDM processors are combined with spatial multiplexing scheme (SMS), as illustrated in Figure \ref{fig:scale-sms}, we up-scale the processor by connecting the photodiodes of parallel WDM processors to combine the output currents. However, this up-scaling reduces the operation speed of the PDs due to the rise in node capacitance resulting from the accumulation of the PD capacitances ($C_{PD}$) along with the wiring capacitance ($C_W$). The wiring capacitance is estimated based on the layout. In the MDM scheme (Figure \ref{fig:scale-mdm}), we combine the output currents in the electro-optic domain by integrating WDM processor instances using MDM. To support more modes in a waveguide, we need to increase the waveguide width and, consequently, the PD width to maintain maximum responsivity. Although this increase in PD width reduces speed due to higher junction capacitance of the multimode PD ($C_{PD}^{MM}$), the gain in the number of supported modes more than offsets this drawback. Additionally, there is no wiring capacitance involved in this approach. The number of operations per second (OPS) for SMS and MDM schemes are calculated using equations \ref{eq:ops_sms} and \ref{eq:ops_mdm} of the manuscript, respectively. It is noteworthy that the ratio between $C_{PD}^{MM}$ and the single-mode PD capacitance, $C_{PD}^{SM}$, is calculated based on the ratio between their respective widths, and the number of supported modes at each waveguide width was extracted from waveguide eigenmode simulations. The OPS versus the number of combined WDM processor instances using SMS and MDM techniques is reported in Figure \ref{fig:scalability} of the manuscript. As we observe, the MDM scheme increases the OPS with up-scaling, as opposed to SMS where the OPS decreases. The MDM scheme offers approximately 4.1 times improvement in OPS as the system up-scales. The parameter values used for this analysis are listed in Table \ref{Tb:scale}.

{
\begin{center}
\begin{tabular}{|c|c|c|c|c|}
\hline 
Parameter &  Value \tabularnewline
\hline 
\hline 
Operation speed of single-mode PD ($F$) & 50 GHz \tabularnewline
\hline
Reference PD capacitance ($C_{PD}$) & 12 pF\tabularnewline
\hline
Wiring capacitance to connect two PDs ($C_W$) & 12 pF\tabularnewline
\hline
\end{tabular}
\end{center}
\captionof{table}{The parameters used to calculate the number of operations per second for SMS and MDM schemes.\label{Tb:scale}}
}

\bibliographystylesupp{sn-nature}
\bibliographysupp{sn-bibliography}

\end{document}